\documentclass[modern]{aastex701}

\usepackage{amssymb}
\usepackage{lipsum}
\usepackage{bm}
\usepackage{amsmath}
\usepackage{wasysym}
\usepackage{cancel}
\usepackage{xspace}
\usepackage{comment}

\newcommand{\pdf}{\mathrm{P}}

\newcommand{\Ntot}{N_\mathrm{tot}}
\newcommand{\Npos}{N_\mathrm{pos}}

\newcommand{\CPP}{\mathbb{C}}
\newcommand{\Cmax}{\CPP_{\mathrm{max}}}

\newcommand{\NtotA}{N_\mathrm{tot,A}}
\newcommand{\NtotB}{N_\mathrm{tot,B}}
\newcommand{\NposA}{N_\mathrm{pos,A}}
\newcommand{\NposB}{N_\mathrm{pos,B}}

\begin{document}


\title{The Catastrophic Consequences of Agnosticism for Life Searches and a Possible Workaround}

\author[orcid=0000-0002-4365-7366,sname='Kipping']{David Kipping}
\affiliation{Columbia University, 550 W 120th Street, New York NY 10027, USA}
\email[show]{dkipping@astro.columbia.edu}

\begin{abstract}
Planned and ongoing searches for life, both biological and technological, confront an epistemic barrier concerning false positives - namely, that we don't know what we don't know. The most defensible and agnostic approach is to adopt diffuse (uninformative) priors, not only for the prevalence of life, but also for the prevalence of confounders. We evaluate the resulting Bayes factors between the null and life hypotheses for an idealized experiment with $\Npos$ positive labels (biosignature detections) among $\Ntot$ targets with various priors. Using diffuse priors, the consequences are catastrophic for life detection, requiring at least ${\sim}10^4$ (for some priors ${\sim}10^{13}$) surveyed targets to ever obtain ``strong evidence'' for life. Accordingly, an HWO-scale survey with $\Ntot{\sim}25$ would have no prospect of achieving this goal. A previously suggested workaround is to forgo the agnostic confounder prior, by asserting some upper limit on it for example, but we find that the results can be highly sensitive to this choice - as well as difficult to justify. Instead, we suggest a novel solution that retains agnosticism: by dividing the sample into two groups for which the prevalence of life differs, but the confounder rate is global. We show that a $\Ntot=24$ survey could expect 24\% of possible outcomes to produce strong life detections with this strategy, rising to $\geq50$\% for $\Ntot\geq76$. However, AB-testing introduces its own unique challenges to survey design, requiring two groups with differing life prevalence rates (ideally greatly so) but a global confounder rate.
\end{abstract}




\section{Introduction}
\label{sec:intro}

The search for life beyond our planetary environment confronts many technical challenges, but one problem raises a potentially severe epistemic barrier - \textit{we don't know what we don't know}.

The remote detection of biology or technology, after centuries of philosophizing, has become a serious goal for 21st century astronomy \citep{schwieterman:2018,fujii:2018,wright:2019}. The 2020 Decadal Survey makes it clear that this is a broad community goal, writing that ``the effort to identify habitable Earth-like worlds in other planetary systems and search for the biochemical signatures of life will play a critical role in determining whether life exists elsewhere in the Universe'' \citep{decadal:2020}. Although that quote focuses on biosignatures, \citet{wright:2022} argue that technosignatures represent an equally promising route to life detection, and thus, in what follows, all references to ``biosignatures'' implicitly encompass technosignatures too.

If detecting a biosignature equated to detecting life, the challenges facing us would be largely engineering ones - primarily a question of building sufficiently large apertures with sufficiently precise instrumentation \citep{rauscher:2016,wang:2018,krissansen:2025}. As appealing as that simplified picture may sound, there has been a growing recognition within the community that biosignatures can also be produced with no life involved; often referred to as false positives or confounding positives \citep{grenfell:2014,rein:2014,reed:2024}. For example, recent work has revealed that substantial concentrations of the O$_2$/O$_3$ biomarker can be produced via at least three different abiotic pathways \citep{selsis:2002,wordsworth:2014,tian:2014} - but presumably more as-yet-unimagined pathways may exist too.

Despite much recent work in this area, the notion of an astrobiological confounder is hardly novel. In 1906, \textit{The New York Times} ran the front page headline ``THERE IS LIFE ON MARS'', in reference to the claims of Martian canals \citep{lowell:1906}, going as far as to state that there was ``absolute proof that there is conscious, intelligent, organic life on Mars'' \citep{nyt:1906}. From a modern perspective, it is easy to regard this optimism as misguided. However, this reflects not a failure of observational skill, but the fact that confounding influences (notably Gestalt reconfiguration) were only just starting to be understood at the time \citep{evans:1903,sternberg:2012}. Lowell (presumably) simply didn't know about Gestalt psychology and, indeed, none of us can know what we don't yet know.

Indeed, there is a long history of claimed signatures of life that were initially presumed to be effectively unique, but then later explained abiotically. As another example, phosphine was proposed as a biosignature by \citet{sousasilva:2020}, one with ``no known abiotic false positives on terrestrial planets''. Shortly after \citet{greaves:2021} presented evidence for ${\sim}$parts-per-billion Venusian phosphine, \citet{bains:2021} reinforced that known abiotic processes struggled to explain the observed levels. However, subsequent work identified multiple pathways, such as volcanogenic phosphide hydrolysis \citep{truong:2021}, radical-mediated phosphate reduction \citep{ferus:2022} and photochemical radical reduction of oxidized phosphorus \citep{mrazikov:2024}.

Similarly, in the domain of technosignatures, narrowband radio signals - particularly those centered near the 21\,cm hydrogen line - have long been regarded as distinctive products of technology, dating back to the seminal SETI proposal of \citet{cocconi:1959}. However, recent work by \citet{mendez:2024} proposed a natural astrophysical mechanism involving stimulated emission from cold hydrogen clouds that can produce narrowband signals near the 1420\,MHz line, potentially explaining events like the 1977 ``Wow! Signal'' without invoking technology.

These lessons pose a significant headache for near-term multi-billion dollar investments to look for life, such as that proposed by HWO, the Habitable Worlds Observatory \citep{decadal:2020,feinberg:2024} or LIFE, the Large Interferometer For Exoplanets \citep{quanz:2022}. We might wonder - are we destined to perpetually repeat this pattern? Initial exuberance of some recently claimed biosignature followed by sober reflection of the myriad of alternative hypotheses. Put another way, can we ever identify a biosignature which exceeds the impressive imagination of theorists to later explain without biology? And, even if so, is that sufficient? Should we not concede that there may simply exist limits to what we can know?

As discussed in \citet{kipping:2024}, it must also be acknowledged that the alien hypothesis is not a typical scientific hypothesis. The alien hypothesis can be invoked to explain almost any anomaly (e.g. \citealt{loeb:2022}), plausibly evade any detection (did you look under \textit{that} rock?) and, as we've already seen, their claims are structurally vulnerable to being rendered obsolete by future natural explanations. Building upon prior Bayesian framings of this problem by \citet{catling:2018} and \citet{walker:2018}, \citet{foote:2023} suggest that there are only two ways out of this predicament: i) a strong prior for life's occurrence, or, ii) a biosignature with no confounders.

Certainly, a strong prior in favor of a specific hypothesis makes that hypothesis favored more readily, a-posteriori. But, this is arguably just Rosenthal bias - experimenters obtaining the very result they expect even when no true signal is present \citep{rosenthal:1963}. The limiting extreme is a Dirac delta prior where no amount of new data can ever influence our beliefs, which means the information gain from the prior to the posterior is formally zero\footnote{
This can be seen by considering the KLD, Kullback-Leibler Divergence \citep{kullback:1951}, between a prior, $\pdf(f)$, and a posterior, $\pdf(f|\mathcal{D})$. The KLD information gain is $\int \pdf(f|\mathcal{D}) \log[\pdf(f|\mathcal{D})/\pdf(f)]\,\mathrm{d}f$, and thus in the limit of dogmatic prior we have $\pdf(f|\mathcal{D}) \to \pdf(f)$ and thus the KLD is zero.
} - what \citet{kipping:2024} call the ``dogmatic experimenter''. Indeed, more generally, stronger priors minimize information gain. On this basis, a reasonable and standard prescription is to adopt a so-called uninformative (although this work favors the more accurate description of ``diffuse'') prior for the existence of life associated with some target. By making the prior minimally informative, the information gain from the experiment is maximized i.e. we learn the most.

Of course, a diffuse prior for life removes one of the two solutions proposed by \citet{foote:2023}. This leaves us with the one hope: an unambiguous biosignature. One suggestion is to measure the assembly index of molecules, for which it is argued that molecules with both high abundance and complexity are unambiguously caused by life \citep{marshall:2021,rutter:2025}. However, such inference requires mass spectroscopy (which HWO will not have). Yet more, one might question, in a philosophical sense, to what degree one can ever really claim certainty about what non-living systems can and cannot do, given that we lack a complete understanding of nature. Although, some technosignatures arguably sit at the apex of what one can defensibly label as a signal with no plausible confounders \citep{wright:2022}. For example, a laser using binary pulses to write out $\pi$ to 100 places would be difficult to conceive of a natural confounder.

However, the kinds of life surveys advocated for \citet{decadal:2020} are a far cry from such a regime. HWO or LIFE will, at best, find spectroscopic signatures indicative of (ideally) multiple gases expected to be produced by a biosphere. And clearly our understanding of the physical, chemical and geological environments of potentially very unfamiliar worlds is evolving - planets are arguably among the most complex astrophysical systems of them all. Even constellations of biosignatures can be reproduced abiotically \citep{rein:2014} and thus we must explicitly concede ignorance in the strict sense: the existence of confounding mechanisms that are presently unknown and perhaps unimagined.

Such a scenario motivates an agnostic approach to confounders. In the same way that we readily acknowledge that we lack a justifiable strong prior for life, we here also concede that we lack a strong prior for confounders. Accordingly, not only do we require a diffuse prior for the prevalence of life (given by $f$), but also for the confounder positive probability, CPP (given by $\CPP$). In this work, we formally evaluate the consequences of such a pairing on our ability to evaluate the evidence for and against the hypothesis of life within some sample (e.g. exoplanets).

\section{Experimental Setup}
\label{sec:setup}

\subsection{An Idealized Experiment}
\label{sub:idealized}

Consider an experiment where the outcome is a simple binary classification: a positive detection ($\mathcal{S}$), or a non-detection, ($\bar{\mathcal{S}}$). Although this could in principle concern any arbitrary phenomenon, in this work we are focussed on the problem of detecting a specific biosignature or technosignature (or a specific constellation of them) produced by extraterrestrial life (or artificial agents derived from life). Note, that for the sake of conciseness, we refer to the signal sought as simply the ``biosignature'' in what follows, although it should be understood to encompass all of the above.

A positive biosignature label, $\mathcal{S}$, can occur via three different pathways: i) the detection of the biosignature that was produced by life, or, ii) the detection (or apparent detection) of the biosignature that was produced abiotically and thus by a confounding factor, or, iii) a false positive generated by the detector itself.

In what follows, we assume the idealized scenario of a perfect detector which never generates false positives. Further, we also assume the detector has a perfect completeness (true positive rate) to the biosignature in question. In other words, if the biosignature is present above some pre-defined threshold, the detector will always successfully recover it. These assumptions are, of course, not practically achievable in a real survey, but they represent the absolute best case scenario. Accordingly, whatever conclusions we derive under such assumptions represent the ceiling of what's practically possible - the real world can only hope to asymptotically approach the results we will report in this work.

\subsection{Confounders}
\label{sub:confounders}

The idealized assumptions made in Section\ref{sub:idealized} greatly simplify our analysis and allow us to isolate the impact of the confounding positive probability, CPP ($\CPP$). The confounder could caused by any number of possible sources, such as unknown/unanticipated chemistry on the target planet, confounding stellar activity, foreground dust contaminations (especially for Dyson sphere searches), etc. It is not our intent to try and predict all possibilities, which we argue is a hopeless task - we cannot know what we cannot know. Instead, our task is simply to accept confounders will always exist and try to agnostically accommodate that fact in our interpretation of said survey.

\subsection{Survey Outcomes}
\label{sub:survey}

Consider that our survey spans a sample of $\Ntot$ targets (e.g. planets) for which one can define that some fraction, $f$, have a life-generated biosignature present\footnote{The true number of inhabited targets could of course be greater than $f$, since some inhabited targets will not manifest the biosignature sought.}. We consider that the primary question that our survey seeks to address it weigh two hypotheses against one another:

\begin{itemize}
\item[{$\mathcal{H}_{f=0}$:}] The life-generated biosignature is not present in any target in the sample i.e. $f=0$.
\item[{$\mathcal{H}_{f>0}$:}] The life-generated biosignature is present in at least some of the targets in the sample i.e. $f>0$.
\end{itemize}

We consider that our survey has three possible outcomes: i) strong evidence for $f=0$ ($\mathcal{H}_{f=0}$), or, ii) strong evidence for $f>0$ ($\mathcal{H}_{f>0}$), or, iii) no conclusive evidence for either hypothesis. We define ``strong evidence'' as a Bayes factor exceeding 10. Accordingly, case i) can be formalized as $\pdf(f=0|\Ntot,\Npos)/\pdf(f>0|\Ntot,\Npos) > 10$ and case ii) is $\pdf(f>0|\Ntot,\Npos)/\pdf(f=0|\Ntot,\Npos) > 10$ (and case iii) spans all other outcomes). Throughout, we treat $\Npos$ as the observed data, while $f$ and $\CPP$ are latent parameters to be marginalized over unless stated otherwise.

\subsection{Likelihood Function}
\label{sub:likelihood}

\begin{figure*}
\begin{center}
\includegraphics[width=8.0cm,angle=0,clip=true]{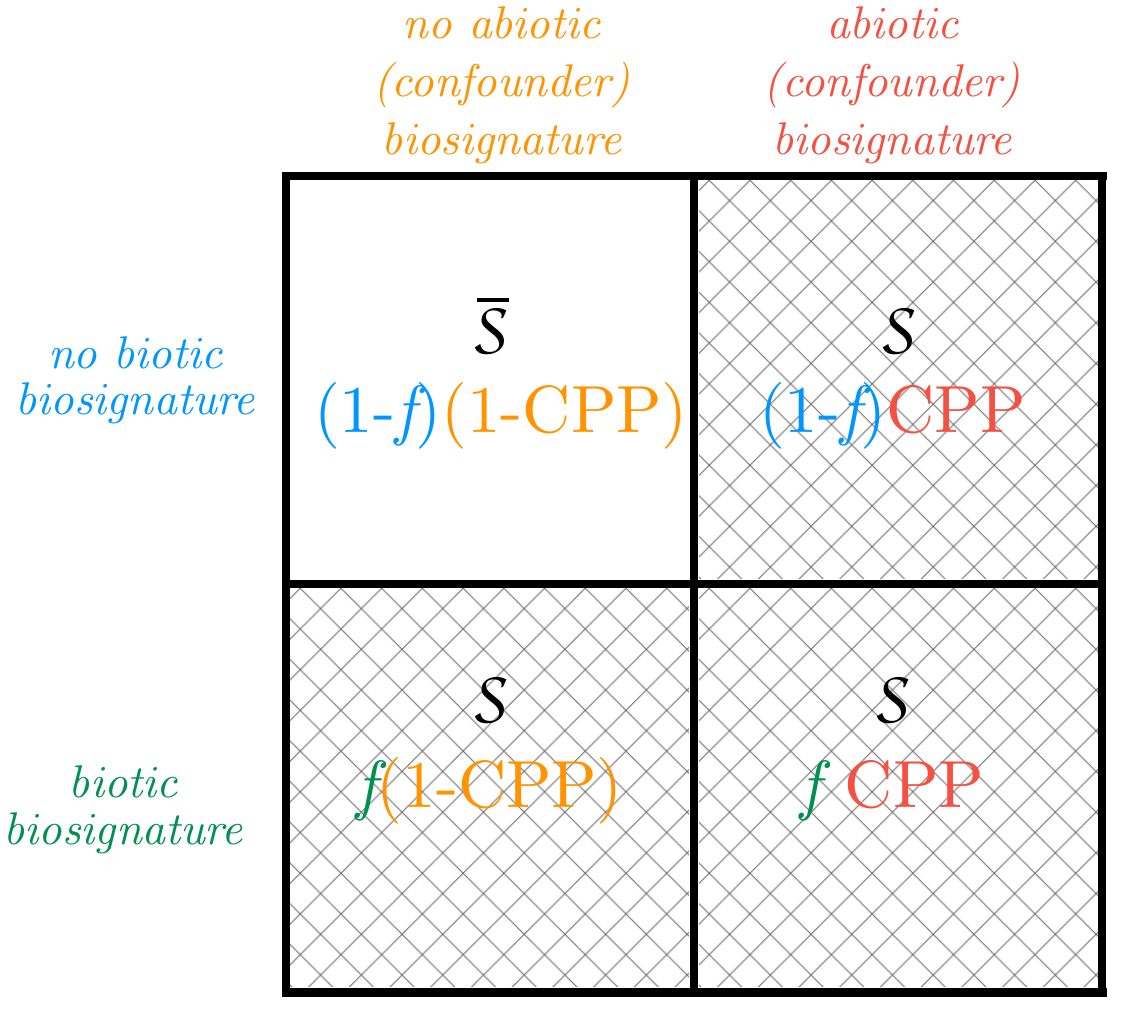}
\caption{
Matrix of possible experimental labels that are assigned in an idealized binary classification experiment. A positive detection label is given by the symbol $\mathcal{S}$, whereas a non-detection label is $\bar{\mathcal{S}}$. A positive detection will be assigned in three of the four possible outcomes. In each quadrant, we present the label and the associated probability.
}
\label{fig:tree}
\end{center}
\end{figure*}

For a single target, a positive detection label ($\mathcal{S}$) can occur via either a life-generated biosignature, or a confounder-generated biosignature, or indeed both acting. The 2x2 matrix of possible outcomes is summarized in Figure~\ref{fig:tree}.

The only way to obtain a non-detection label ($\bar{\mathcal{S}}$) is if neither act, which has a probability of $(1-f) \times (1-\CPP)$. Thus, the probability of a positive detection must be $1 - (1-f) (1-\CPP)$. Therefore, for a survey yielding $\Npos$ positive detections from a sample of $\Ntot$ targets, the likelihood function ($\mathcal{L}$) must follow a binomial distribution, given by

\begin{align}
\underbrace{\pdf(\Npos|f,\CPP,\Ntot)}_{=\mathcal{L}} &=
\big( 1 - (1-f)(1-\CPP) \big)^{\Npos}
\big( (1-f)(1-\CPP) \big)^{\Ntot-\Npos}
\binom{\Ntot}{\Npos}.
\label{eqn:likelihood}
\end{align}

\section{Limiting Case of a Known Confounder Positive Probability}
\label{sec:knownCPP}

\subsection{General $\Npos$ case}

In general, the confounding positive rate is unknown - especially since it may depend upon the prevalence of unanticipated physical mechanisms. However, for the sake of illustrating its role more clearly, we initially consider the case where $\CPP$ is a known, fixed scalar. For $f$, we adopt a Beta distribution prior where the mean is fixed to $\tfrac{1}{2}$, such that

\begin{align}
\pdf(f|\alpha) &= \frac{ ( f (1-f) )^{\alpha-1} }{ B[\alpha,\alpha] },
\label{eqn:betapdf}
\end{align}

where $\alpha$ governs the shape of the prior. For example, setting $\alpha=1$ causes the prior to equal a uniform distribution, $\mathcal{U}[0,1]$. Although a uniform prior is intuitively attractive as a non-informative prior for $f$, the Jeffrey's prior for a binomial process such as this is found by setting $\alpha=1/2$, which we denote as $\mathcal{J}[0,1]$. One may write that the posterior distribution of $f$ is:

\begin{align}
\underbrace{ \pdf(f|\Ntot,\Npos,\CPP,\alpha) }_{=\mathcal{P}_{\mathcal{B}}} 
&= \frac{ \pdf(\Npos|f,\Ntot,\CPP) \pdf(f|\alpha) }{ \underbrace{\pdf(\Npos|\Ntot)}_{=\mathcal{Z}_{\mathcal{B}}} },
\end{align}

where $\mathcal{Z}$ is the marginal likelihood necessary to normalize the posterior distribution, where we add the subscript ``$\mathcal{B}$'' to denote this is for the case of a Beta prior. The marginal likelihood, also known as the Bayesian evidence, is found by integration\footnote{Throughout this paper, several intermediate results are stated without derivation; these integrals admit closed-form solutions obtainable symbolically (e.g. via standard computer algebra systems), and are verified numerically where appropriate.} of the likelihood multiplied by the prior.

\begin{align}
\mathcal{Z}_{\mathcal{B}} &=
\int_{f=0}^1
\pdf(\Npos|f,\Ntot,\CPP) \pdf(f|\alpha)\,\mathrm{d}f,
\end{align}

which, using Equations~\ref{eqn:likelihood} \& \ref{eqn:betapdf}, evaluates to

\begin{align}
\mathcal{Z}_{\mathcal{B}} = 
\Bigg(&\frac{
\Gamma[\alpha] \Gamma[\Ntot-\Npos+\alpha]
\CPP^{\Npos} (1-\CPP)^{\Ntot-\Npos} \binom{\Ntot}{\Npos} }{
B(\alpha,\alpha) }\Bigg) \times \nonumber\\
\qquad \Bigg(&
(\Ntot+2\alpha) \,_2\tilde{F}_1[
-\Npos,\alpha;\Ntot-\Npos+1+2\alpha;\frac{\CPP-1}{\CPP}
] - \nonumber\\
\qquad& \Npos \, _2\tilde{F}_1[
1-\Npos,\alpha;\Ntot-\Npos+1+2\alpha;\frac{\CPP-1}{\CPP}
]
\Bigg),
\label{eqn:ZB}
\end{align}

where $\Gamma[x]$ is the Gamma function and $_2\tilde{F}_1$ is the regularized hypergeometric function. The appearance of hypergeometric functions here is unavoidable: when the $\CPP$ is fixed, the marginal likelihood reduces to a single-parameter binomial model in $f$, whose normalization integrals correspond to standard Euler-type integrals with closed forms in terms of $_2\tilde{F}_1$.

This result in Equation~\ref{eqn:ZB} is one of the two components necessary to answer the primary question of the survey: does the data favor $\mathcal{H}_{f=0}$ or $\mathcal{H}_{f>0}$? To address this, we turn to the Bayes factor, $\mathcal{K}$. In general, the probabiliy of obtaining some data, $\mathcal{D}$, given a hypothesis $\mathcal{H}_1$, versus some other hypothesis $\mathcal{H}_2$, is

\begin{align}
\frac{\pdf(D|\mathcal{H}_1)}{\pdf(D|\mathcal{H}_2)}
\end{align}

Equation~\ref{eqn:ZB} yields $\pdf(D|\mathcal{H}_{f>0})$, and thus to complete the picture, we need to know $\pdf(\Npos|f=0,\Ntot,\CPP)$ - the likelihood of obtaining the data ($\Npos$) when $f=0$. In this limit, the likelihood function (Equation~\ref{eqn:likelihood}) simplifies to

\begin{align}
\mathcal{L}_{f=0} \equiv \lim_{f\to0} \mathcal{L} &=
\big( \CPP \big)^{\Npos}
\big( 1-\CPP \big)^{\Ntot-\Npos}
\binom{\Ntot}{\Npos},
\end{align}

Hence, the Bayes factor of $\mathcal{H}_{f=0}$ relative to $\mathcal{H}_{f>0}$ is

\begin{align}
\mathcal{K}_{\mathcal{B}}
&= \frac{ \pdf(\mathcal{D}|\mathcal{H}_{f=0}) }{ \pdf(\mathcal{D}|\mathcal{H}_{f>0}) }
 = \frac{ \mathcal{L}_{f=0} }{ \mathcal{Z}_{\mathcal{B}} }.
\label{eqn:KB}
\end{align}

In the limit of $\alpha\to1$, where the prior on $f$ becomes uniform (see Equation~\ref{eqn:betapdf}), this becomes

\begin{align}
\mathcal{K}_{\mathcal{U}} &= \frac{ \Ntot-\Npos+1 }{
_2F_1[ 1,-\Npos;\Ntot-\Npos+2;\frac{\CPP-1}{\CPP} ]
}.
\label{eqn:KU}
\end{align}

And, in the case of $\alpha\to1/2$, where the prior becomes a Jeffrey's prior (see Equation~\ref{eqn:betapdf}), the Bayes factor becomes

\begin{align}
\mathcal{K}_{\mathcal{J}} &= \frac{ \sqrt{\pi} }{ \Gamma[\Ntot-\Npos+\tfrac{1}{2}]
_2\tilde{F}_1[
\tfrac{1}{2},-\Npos;\Ntot-\Npos+1;\frac{\CPP-1}{\CPP}
]
}.
\label{eqn:KJ}
\end{align}

\begin{figure*}
\begin{center}
\includegraphics[width=16.0cm,angle=0,clip=true]{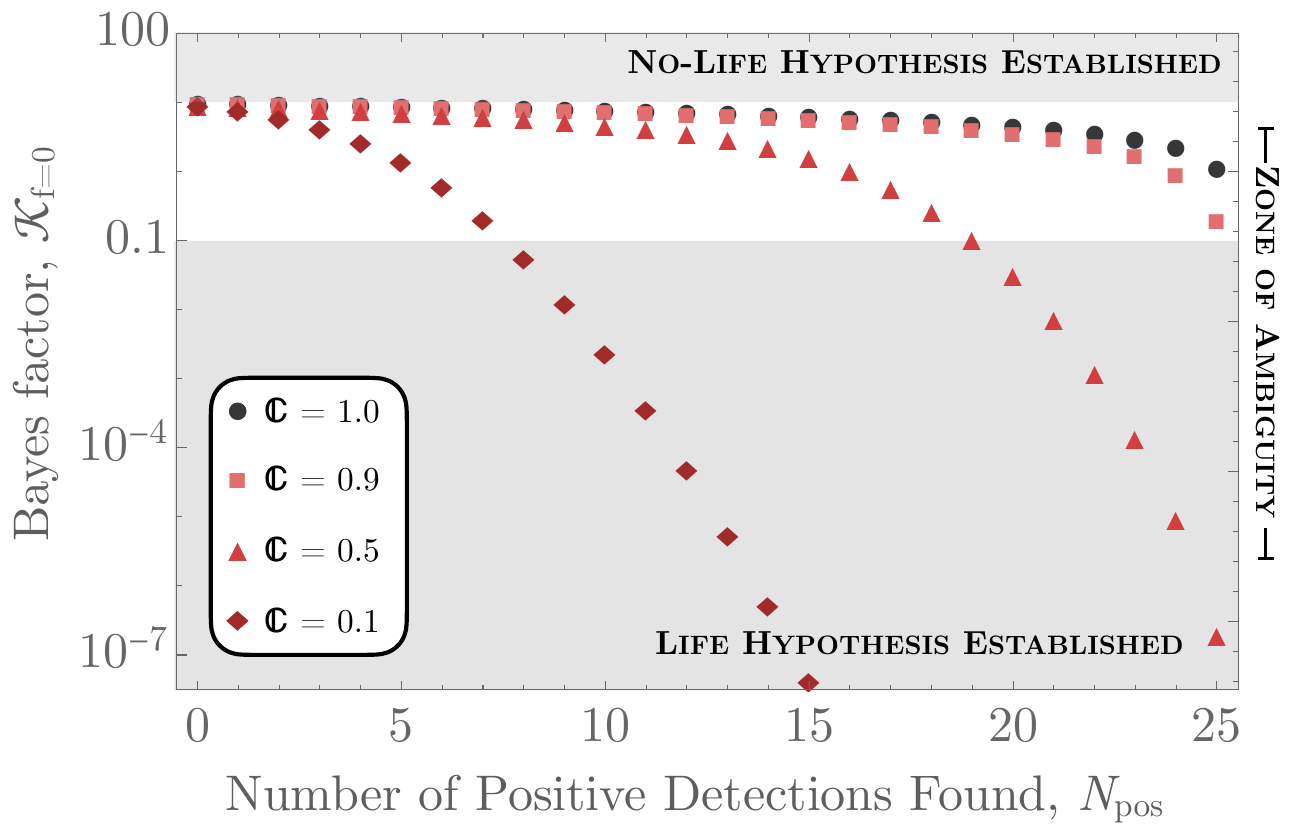}
\caption{
Assuming a known confounder rate ($\CPP$), we here show the Bayes factor of the null hypothesis that life is absent, $f=0$, versus the affirmative hypothesis that life is present, $f>0$, as a function of the number of positive detection labels obtained by an HWO-like survey ($\Ntot=25$). The four different markers correspond to four different choices for $\CPP$ - here unrealistically treated as a perfectly known fixed scalar. The figure reveals that the null hypothesis is impossible to ever establish here, but establishing the affirmative hypothesis is possible when $\CPP\leq0.87$, and becomes increasingly easier as $\CPP$ decreases.
}
\label{fig:fixedCPP}
\end{center}
\end{figure*}

To provide some intuition, we plot the resulting Bayes factors, under the assumption $f \sim \mathcal{J}[0,1]$, for various choices of $\CPP$ in Figure~\ref{fig:fixedCPP}.

\subsection{Special case of $\Npos=0$}

The strongest possible evidence for hypothesis $\mathcal{H}_{f=0}$ will occur when $\Npos=0$. Accordingly, it is instructive to evaluate the Bayes factor in the limit to see if ``strong evidence'' ($\mathcal{K}>10$) can be achieved. If it cannot be achieved in this limit, then $\Npos>0$ will certainly not yield strong evidence either. For the general Beta prior, we find

\begin{align}
\lim_{\Npos\to0} \mathcal{K}_{\mathcal{B}} &= \frac{
B[\alpha,\alpha]\Gamma[\Ntot+1+2\alpha]
}{
(\Ntot+2)\Gamma[\alpha]\Gamma[\Ntot+\alpha]
}.
\label{eqn:KB_Npos0}
\end{align}

In the case of $\alpha\to1$ (uniform prior on $f$), we find that $\Ntot\geq9$ is required for a strong detection in this limit. In the case of $\alpha\to1$ (Jeffrey's prior on $f$), this becomes $\Ntot\geq32$.

A nominal sample size for HWO is $\Ntot=25$ \citep{feinberg:2024,stark:2024} and thus already this sample size can be challenged as being hardly sufficient to strongly support the null hypothesis even in the limit of an ideal experiment and a Jeffrey's prior on biosignature occurrence.

\subsection{Special case of $\Npos=\Ntot$}

The other extreme is the limit of $\Npos=\Ntot$, where one obtains the strongest evidence possible for $\mathcal{H}_{f>0}$ i.e. we minimize $\mathcal{K}$, or equivalently maximize $1/\mathcal{K}$. Even if $\Ntot$ is sufficient to establish the null hypothesis when $\Npos=0$, that says nothing about whether we can establish the affirmative hypothesis. And, as before, if we cannot establish the affirmative hypothesis in the limit of $\Npos=\Ntot$, then there is no hope of doing so when $\Npos<\Ntot$. In this way, the following result provides a firm boundary on what constitutes a (potentially) useful experiment. Taking the limit $\Npos\to\Ntot$ on the Bayes factor for $\alpha=1$, we find

\begin{align}
\lim_{\Npos\to\Ntot} \mathcal{K}_{\mathcal{U}} &= \frac{
(1-\CPP) \CPP^{\Ntot} (\Ntot+1)
}{
1 - \CPP^{\Ntot+1}
}.
\label{eqn:KU_NposA}
\end{align}

For an HWO-like sample size of $\Ntot=25$, the above indicates that establishing the affirmative hypothesis is only possible if $\CPP<0.867$. In the case of $\alpha\to\frac{1}{2}$, we find 

\begin{align}
\lim_{\Npos\to\Ntot} \mathcal{K}_{\mathcal{J}} &= \frac{
1
}{
_2\tilde{F}_1[
\tfrac{1}{2},-\Ntot;1;\frac{\CPP-1}{\CPP}
]
}.
\label{eqn:KJ_NposA}
\end{align}

For an HWO-like sample size of $\Ntot=25$, the above indicates that establishing the affirmative hypothesis is only possible if $\CPP<0.874$. Therefore, in both limits, the Jeffrey's prior provides a more conservative, demanding set of requirements on the experimental design.

Although the above results are instructive, it is difficult to imagine a scenario where the confounding positive rate is known so well it can be treated as a fixed scalar like this. It is formally equivalent to a delta function prior on $\CPP$, infinitesimal uncertainty and perfect precision. In reality, $\CPP$ will have some (non-delta) prior probability distribution, and we explore the consequences of this in the next section.

\section{General Case of a Diffuse Confounder Prior}
\label{sec:diffuse}

In the more realistic case where $\CPP$ has some continuous prior distribution, our first task is to select an appropriate prior. We argue that the most agnostic and general choice is a non-informative prior, or more accurately a ``diffuse'' prior since all priors inform the posterior to some degree. Since we assume here that some confounders will be previously unanticipated, perhaps even unknown, physical mechanisms, one cannot know anything about their prevalence a-priori. In such a scenario, the only path forward is a diffuse prior.

Before continuing, we emphasize that the distribution of outcomes (i.e. $\Npos$) will be highly sensitive to the choice of prior on both $f$ and $\CPP$. This section begins by exploring the most common choice of simple uniform priors on $f$ and $\CPP$, but forewarn the reader that this leads a non-uniform $\Npos$ distribution, something we tackle directly later in Section~\ref{sub:GG}.

\subsection{Uniform-Uniform Prior}
\label{sub:UU}
\subsubsection{General $\Npos$ case}

In this sub-section, we start with the simplifying assumption of a uniform distribution on $\CPP$ i.e. $\CPP \sim \mathcal{U}[0,1]$. Similarly, $f$ is also typically unknown a-priori and thus we also adopt a uniform prior here over the interval $[0,1]$. We acknowledge that a uniform prior is not the Jeffrey's prior of a Bernoulli process and places much more weight at intermediate $f$ values than such a prior. Nevertheless, it is an instructive starting place thanks to its intuitive interpretation and mathematical simplicity. With these priors, we can write the joint posterior distribution of $\{f,\CPP\}$ as:

\begin{align}
\underbrace{ \pdf(f,\CPP|\Ntot,\Npos) }_{=\mathcal{P}_{\mathcal{U}\mathcal{U}}} 
&= \frac{ \pdf(\Npos|\Ntot,f,\CPP) \pdf(f) \pdf(\CPP) }{ \underbrace{\pdf(\Npos|\Ntot)}_{=\mathcal{Z}_{\mathcal{U}\mathcal{U}}} },
\end{align}

where we add the subscript ``$\mathcal{U}\mathcal{U}$'' to denote this is for the case of a double uniform prior. The marginal likelihood is found by double integration of the likelihood multiplied by the prior:

\begin{align}
\mathcal{Z}_{\mathcal{U}\mathcal{U}} &= \int_{\CPP=0}^1 \int_{f=0}^1
\big( 1 - (1-f)(1-\CPP) \big)^{\Npos}
\big( (1-f)(1-\CPP) \big)^{\Ntot-\Npos}
\binom{\Ntot}{\Npos}\,\mathrm{d}f\,\mathrm{d}\CPP,
\end{align}

which evaluates to

\begin{align}
\mathcal{Z}_{\mathcal{U}\mathcal{U}} &= \frac{ H_{\Ntot+1} - H_{\Ntot-\Npos} }{ \Ntot+1 },
\label{eqn:ZUU}
\end{align}

where $H_n$ is the $n^{\mathrm{th}}$ harmonic number. The harmonic numbers arise from summing inverse linear factors generated by the marginalization over the binomial success probability, reflecting the logarithmic volume of parameter space compatible with a given number of positives.

As before, this result is just one of the two components necessary to answer the primary question of the survey: does the data favor $\mathcal{H}_{f=0}$ or $\mathcal{H}_{f>0}$? Next, we need the marginal likelihood in the limit of $f\to0$:

\begin{align}
\mathcal{Z}_{f=0,\mathcal{U}\mathcal{U}} &= \int_{\CPP=0}^1
\mathcal{L}_{f=0}
\,\mathrm{d}\CPP,\\
\qquad&= \frac{1}{\Ntot+1}.
\label{eqn:Zf0UU}
\end{align}

Equipped with $\mathcal{Z}_{\mathcal{U}\mathcal{U}}$ and $\mathcal{Z}_{f=0,\mathcal{U}\mathcal{U}}$, we can write that the Bayes factor that $f=0$ versus some general $f$ value (i.e. $f>0$) is

\begin{align}
\mathcal{K}_{\mathcal{U}\mathcal{U}} =& \frac{ \mathcal{Z}_{f=0,\mathcal{U}\mathcal{U}} }{ \mathcal{Z}_{\mathcal{U}\mathcal{U}} },\nonumber\\
\qquad&= \frac{1}{ H_{\Ntot+1} - H_{\Ntot-\Npos} }.
\label{eqn:KUU}
\end{align}

\begin{figure*}
\begin{center}
\includegraphics[width=16.0cm,angle=0,clip=true]{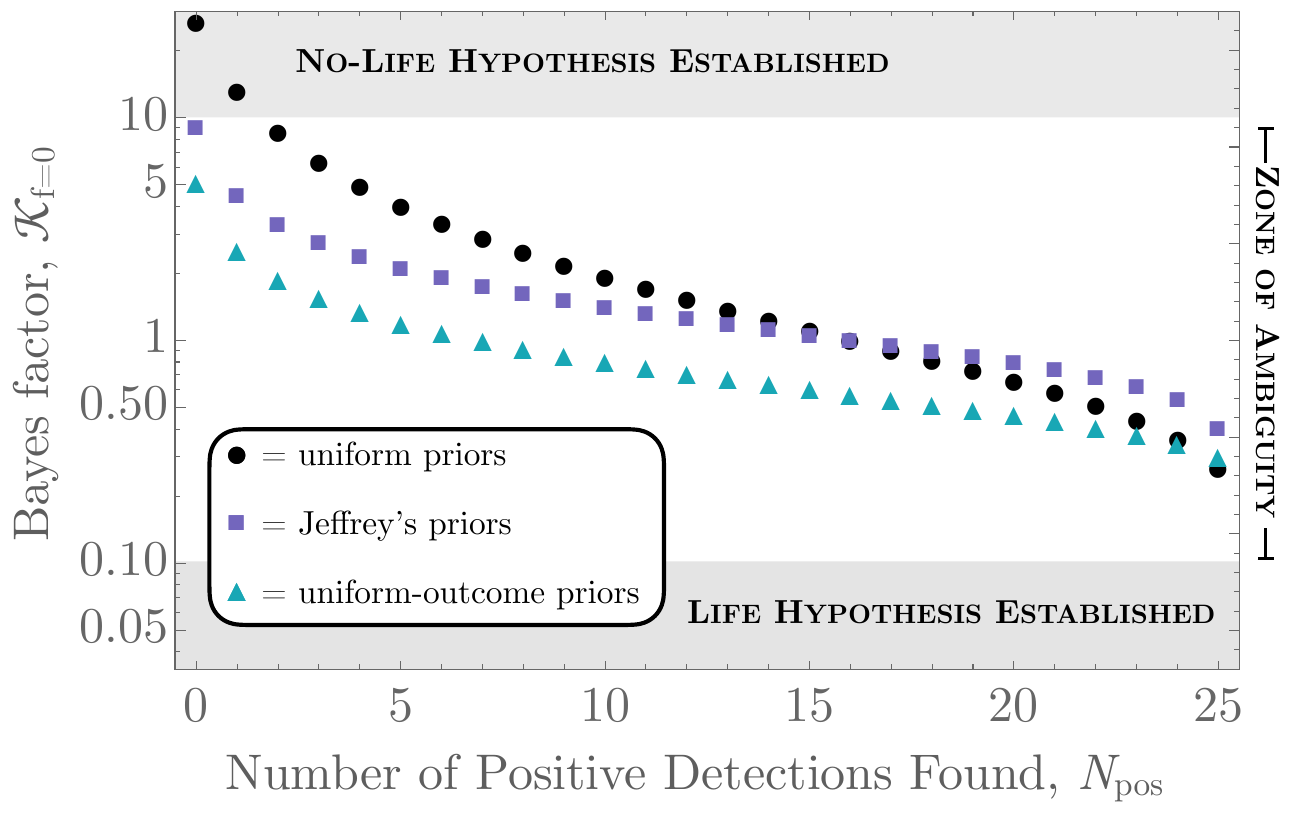}
\caption{
Assuming a diffuse prior for the confounder rate ($\CPP$), we here show the Bayes factor of the null hypothesis that life is absent, $f=0$, versus the affirmative hypothesis that life is present, $f>0$, as a function of the number of positive detection labels obtained by an HWO-like survey ($\Ntot=25$). The three different markers correspond to three different choices for the priors on $f$ and $\CPP$. The figure reveals that strong evidence for the life hypothesis is impossible to ever obtain under these priors.
}
\label{fig:diffuse}
\end{center}
\end{figure*}

Figure~\ref{fig:diffuse} evaluates $\mathcal{K}_{f=0,\mathcal{U}\mathcal{U}}$ assuming $\Ntot=25$, which is fiducial sample size sought by HWO, and all possible $\Npos$ realizations. Inspection Figure~\ref{fig:diffuse} confirms two of our previous assertions. First, the strongest evidence for $f=0$ occurs if $\Npos=0$. In other words, obtaining no detections whatsoever maximally strengths the hypothesis that $f=0$, as expected. The Bayes factor in favor of $f=0$ then smoothly drops monotonically to the extreme case of $\Npos=\Ntot$, where one obtains the strongest evidence for $f>0$ (since $1/\mathcal{K}_{f=0}$ is maximized here).

\subsubsection{Special case of $\Npos=0$}

Our inspection of Figure~\ref{fig:diffuse} reveals that all of the Bayes factors are bound between the two extreme end cases of $\Npos=0$ and $\Npos=\Ntot$, such that

\begin{align}
\lim_{\Npos\to\Ntot}\mathcal{K} \leq \mathcal{K} \leq \lim_{\Npos\to0}\mathcal{K}.
\end{align}

This behaviour is intuitive: observing no positives provides the strongest support for the null hypothesis $f=0$, whilst observing positives in every trial makes the null maximally implausible, with intermediate outcomes interpolating smoothly between these extremes. In the limit of $\Npos=0$, Equation~(\ref{eqn:KUU}) simplifies to

\begin{align}
\lim_{\Npos\to0} \mathcal{K}_{\mathcal{U}\mathcal{U}} &= \Ntot+1.
\label{eqn:KUU_Npos0}
\end{align}

Thus, we obtain the identical result as found the same limiting case in Section~\ref{sec:knownCPP} with a fixed $\CPP$ (see Equation~\ref{eqn:KB_Npos0}). We note that the case considered here, a series of null detections, is the same case considered by \citet{angerhausen:2025}.

Before we continue to the opposite limit of $\Npos=\Ntot$, we briefly note that with the uniform priors assumed here, $\Npos=0$ is a highly improbable outcome a-priori. For example, from $10^5$ Monte Carlo samples, we obtain just 136 draws of $\Npos=0$ using $\Ntot=25$ (a 0.136\% probability outcome). The fact these priors so strongly disfavor $\Npos=0$ is a legitimate concern that we address later.

\subsubsection{Special case of $\Npos=\Ntot$}

In the limit of $\Npos=\Ntot$, the full evidence becomes

\begin{align}
\lim_{\Npos\to\Ntot} \mathcal{K}_{\mathcal{U}\mathcal{U}} &= \frac{1}{H_{\Ntot+1}}.
\label{eqn:KUU_NposA}
\end{align}

To reach a critical threshold of a Bayes factor of 10 in favor of $f>0$, we would thus need $\Ntot=12366$ observations (or more). In comparison to the $\Npos=0$ limit, where just $\Ntot=9$ measurements were needed to strongly favor the null result of $f=0$, this result is quite startling. It is three orders-of-magnitude more difficult to establish the positive existence of a phenomenon than its absence, at least in comparing these two extreme cases.

This can be understood as a product of the unknown confounding positive rate - since even a long series of positive detections could always plausibly be just produced by the perfectly plausible scenario of $\CPP\to1$. Further, $\CPP=1$ is just as likely to explain $\Npos=\Ntot$ as $f=1$ is, since both have uniform priors and influence the likelihood in equal measures. The 12366 detections case corresponds to some mix of the two effects where maintaining $f=0$ essentially becomes statistically untenable.

\subsubsection{Toy Example}

As a simple example of the above applied to a real survey, \citet{suazo:2024} took a sample of $\Ntot{\sim}320000$ infrared sources and compared to spectral energy distribution models from Dyson spheres, finding $\Npos{\sim}11000$ were broadly consistent. \citet{suazo:2024} then performed a battery of tests to thin these down to just seven, but even using $\Npos=11000$ and $\Ntot=320000$ and adopting uniform priors for $f$ and $\CPP$ implies (using our Equation~\ref{eqn:KUU}) strong evidence for that these 11000 positive labels are likely confounders - which is of course what \citet{suazo:2024} concluded independently.


\subsection{Jeffreys-Jeffreys Prior}
\label{sub:JJ}
\subsubsection{General $\Npos$ case}

We next consider the case where $f$ and $\CPP$ follow the Jeffrey's prior for a Bernoulli process. Recall that a Bernoulli process is simply a binary outcome defined by some probability, like a coin toss. Here, $f$ and $\CPP$ represent the probability of a success from such a Bernoulli experiment. A Jeffrey's prior ensures that the distribution is invariant under change of coordinates and is proportional to the square root of the determinant of the Fisher information matrix. In the case of a Bernoulli process, the prior becomes

\begin{align}
\mathcal{J}(x) &= \frac{1}{\pi\sqrt{x(1-x)}}.
\end{align}

Adopting $f\sim\mathcal{J}[0,1]$ and $\CPP\sim\mathcal{J}[0,1]$, we can write the joint posterior distribution of $\{f,\CPP\}$ as:

\begin{align}
\underbrace{ \pdf(f,\CPP|\Ntot,\Npos) }_{=\mathcal{P}_{\mathcal{J}\mathcal{J}}} 
&= \frac{ \pdf(\Npos|\Ntot,f,\CPP) \pdf(f) \pdf(\CPP) }{ \underbrace{\pdf(\Npos|\Ntot)}_{=\mathcal{Z}_{\mathcal{J}\mathcal{J}}} },
\end{align}

where we add the subscript ``$\mathcal{J}\mathcal{J}$'' to denote this is for the case of a double Jeffrey's prior. The marginal likelihood is found by double integration of the likelihood multiplied by the prior:

\begin{align}
\mathcal{Z}_{\mathcal{J}\mathcal{J}} &= \int_{\CPP=0}^1 \int_{f=0}^1
\mathcal{L}(\Npos) \mathcal{J}(f) \mathcal{J}(\CPP)\,\mathrm{d}f\,\mathrm{d}\CPP,
\label{eqn:ZJJ}
\end{align}

for which we are unable to find a closed-form solution, but can be integrated numerically. The marginal likelihood in the limit of $f\to0$ is more tractable, yielding:

\begin{align}
\mathcal{Z}_{f=0,\mathcal{J}\mathcal{J}} &= \int_{\CPP=0}^1
\mathcal{L}_{f=0} \mathcal{J}(\CPP)
\,\mathrm{d}\CPP,\\
\qquad&= \frac{
\binom{\Ntot}{\Npos} \Gamma[\Npos+\tfrac{1}{2}] \Gamma[\Ntot-\Npos+\tfrac{1}{2}]
}{
\pi \Ntot!
}.
\label{eqn:Zf0JJ}
\end{align}

Figure~\ref{fig:diffuse} evaluates $\mathcal{K}_{f=0,\mathcal{J}\mathcal{J}}$ assuming $\Ntot=25$, which is fiducial sample size sought by HWO, and all possible $\Npos$ realizations.

\subsubsection{Special case of $\Npos=0$}

Although we cannot find a closed-form for $\mathcal{Z}_{\mathcal{J}\mathcal{J}}$ for arbitrary $\Npos$, the special case of $\Npos=0$ is much simpler and yields

\begin{align}
\lim_{\Npos\to0} \mathcal{Z}_{\mathcal{J}\mathcal{J}} &= \frac{1}{\pi} \Big(\frac{\Gamma[\Ntot+\tfrac{1}{2}]}{\Gamma[\Ntot+1]}\Big)^2.
\label{eqn:ZJJ_Npos0}
\end{align}

Combining with Equation~\ref{eqn:Zf0JJ} yields a Bayes factor of

\begin{align}
\lim_{\Npos\to0} \mathcal{K}_{\mathcal{J}\mathcal{J}} &= \frac{\sqrt{\pi}\Ntot!}{\Gamma[\Ntot+\tfrac{1}{2}]}.
\label{eqn:KJJ_Npos0}
\end{align}

Thus, we obtain the identical result as found the same limiting case in Section~\ref{sec:knownCPP} with a fixed $\CPP$ (see Equation~\ref{eqn:KB_Npos0}).

\subsubsection{Special case of $\Npos=\Ntot$}

In the limit of $\Npos=\Ntot$, the full evidence integral becomes

\begin{align}
\lim_{\Npos\to\Ntot} \mathcal{Z}_{\mathcal{J}\mathcal{J}} &= 
\int_{\CPP=0}^1 \int_{f=0}^1 \frac{
(1-(1-f)(1-\CPP)^{\Ntot})
}{
\pi^2\sqrt{f(1-f)\CPP(1-\CPP)}
}\,\mathrm{d}f\,\mathrm{d}\CPP.
\label{eqn:ZJJ_NposA}
\end{align}

To make progress, consider the limit of $\Ntot \gg 1$, where numerical testing proves the strong evidence threshold must lie. For the sake of conciseness, we provide the derivation for the Bayes factor under this limit in Appendix A, which proves that

\begin{align}
\lim_{\Ntot\gg1} \lim_{\Npos\to\Ntot} \mathcal{K}_{\mathcal{J}\mathcal{J}} &\simeq \frac{\pi}{\log\Ntot}.
\end{align}

Thus, to reach a Bayes factor of $1/\mathcal{K}\geq10$, we require $\Ntot \gtrsim e^{10\pi} = 4.403 \times 10^{13}$. Accordingly, some 44 trillion observations are necessary, all of which are positive detections, to lead to strong evidence in favor of $\mathcal{H}_{f>0}$. The absurdity of this value (as well as the double uniform case) demonstrates the implausibility of ever establishing the life hypothesis with a diffuse CPP prior.


\subsection{Uniform-Outcomes Priors}
\label{sub:GG}
\subsubsection{Motivation}

\begin{figure*}
\begin{center}
\includegraphics[width=10.0cm,angle=0,clip=true]{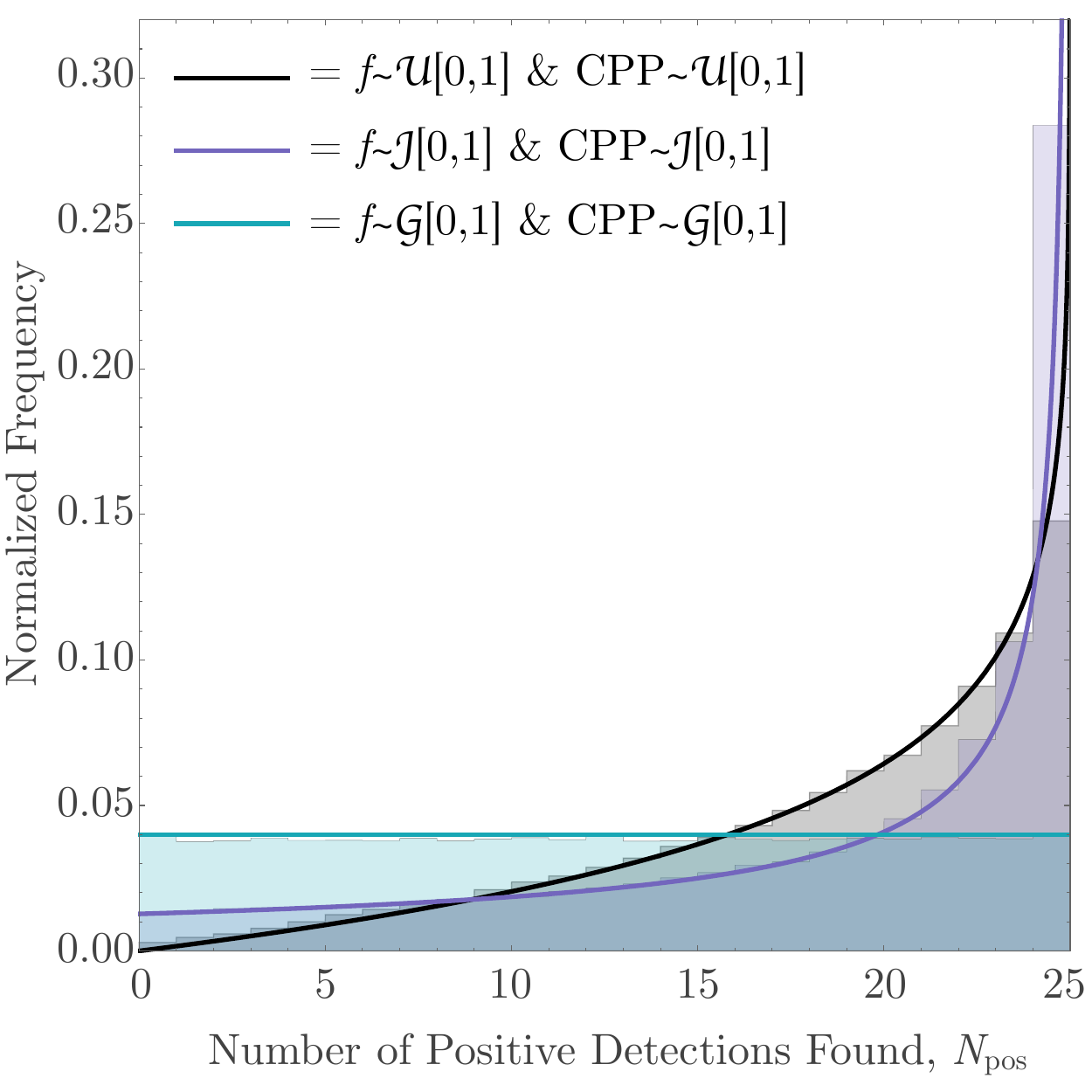}
\caption{
Probability distribution of $\Npos$, the number of positive detection labels that manifest under three different choices for the prior distribution of $f$ and $\CPP$. One might expect that the double-uniform and double-Jeffrey's priors are ideal diffuse priors, but as one can see here they produce a strong bias towards large $\Npos$ values. The final prior, $\mathcal{G}[0,1]$, is engineered to produce a uniform distribution in $\Npos$.
}
\label{fig:Nposdist}
\end{center}
\end{figure*}

In Sections~\ref{sub:UU} \& \ref{sub:JJ}, we evaluated the Bayes factor of the hypotheses $\mathcal{H}_{f=0}$ to $\mathcal{H}_{f>0}$ under two apparently sensible choices for a diffuse prior: a uniform and Jeffrey's prior. However, as noted earlier, such a choice does not produce a uniform distribution in $\Npos$ - the experimental outcome. This is illustrated in Figure~\ref{fig:Nposdist}, where for $\Ntot=25$ we show the resulting histogram from both $10^5$ Monte Carlo draws (histograms) and analytic transformed distributions (smooth lines). This highlights just how biased the uniform ($\mathcal{U}[0,1]$) and Jeffrey's ($\mathcal{J}[0,1]$) priors are, when viewed in the $\Npos$ landscape.

To remedy this, we sought to derive a new prior, $\mathcal{G}(x)$, that could be applied to both $f$ and $\CPP$ and yet guarantee a uniform distribution in $\Npos$. For the sake of brevity, we outline the derivation of this prior in Appendix~B, yielding the result of

\begin{align}
\mathcal{G}(x) &= \frac{1}{ \sqrt{\pi} \sqrt{-\log(1-x)} }.
\label{eqn:Gprior}
\end{align}

We refer to this as the ``uniform-outcomes'' prior in what follows. Imposing that $f \sim \mathcal{G}[0,1]$ and $\CPP \sim \mathcal{G}[0,1]$ ensures a uniform distribution in $p$ and thus $\Npos$. We verified this through numerical testing, as shown in Figure~\ref{fig:Nposdist}.

\subsubsection{General $\Npos$ case}

We now proceed to derive the marginal posteriors and likelihoods as we did before, but now assuming that $f$ and $\CPP$ follow the uniform-outcome's prior. Adopting $f\sim\mathcal{G}[0,1]$ and $\CPP\sim\mathcal{G}[0,1]$, we can write the joint posterior distribution of $\{f,\CPP\}$ as:

\begin{align}
\underbrace{ \pdf(f,\CPP|\Ntot,\Npos) }_{=\mathcal{P}_{\mathcal{G}\mathcal{G}}} 
&= \frac{ \pdf(\Npos|\Ntot,f,\CPP) \pdf(f) \pdf(\CPP) }{ \underbrace{\pdf(\Npos|\Ntot)}_{=\mathcal{Z}_{\mathcal{G}\mathcal{G}}} },
\end{align}

where we add the subscript ``$\mathcal{G}\mathcal{G}$'' to denote this is for the case of a double uniform-outcomes prior. The marginal likelihood is found by double integration of the likelihood multiplied by the prior:

\begin{align}
\mathcal{Z}_{\mathcal{G}\mathcal{G}} &= \int_{\CPP=0}^1 \int_{f=0}^1
\mathcal{L}(\Npos) \mathcal{G}(f) \mathcal{G}(\CPP)\,\mathrm{d}f\,\mathrm{d}\CPP.
\end{align}

The integral can be simplified by subsituting $\sqrt{u} = \sqrt{-\log\bar{f}}$ and $\sqrt{v} = \sqrt{-\log\bar{\CPP}}$, such that $\mathrm{d}\bar{f}=e^{-u}$ and $\mathrm{d}\bar{\CPP}=e^{-v}$ and yielding

\begin{align}
\mathcal{Z}_{\mathcal{G}\mathcal{G}} &= \binom{\Ntot}{\Npos} \int_{v=0}^{\infty} \int_{u=0}^{\infty}
\frac{
e^{-(\Ntot-\Npos+1)(u+v)} (1-e^{-(u+v)})^{\Npos}
}{
\pi\sqrt{u v}
}\,\mathrm{d}u\,\mathrm{d}v.
\end{align}

In this form, one can see that the substitution $s=u+v$ can simplify things further, where we can exploit the fact that

\begin{align}
\int_{v=0}^{\infty} \int_{u=0}^{\infty} \frac{f(u+v)}{\sqrt{u v}}\,\mathrm{d}u\,\mathrm{d}v &= \pi \int_{s=0}^{\infty} f(s)\,\mathrm{d}s,
\end{align}

since $\int_{u=0}^s \mathrm{d}u/\sqrt{u(s-u)} = \pi$. The marginal likelihood is now simply

\begin{align}
\mathcal{Z}_{\mathcal{G}\mathcal{G}} &= \binom{\Ntot}{\Npos} \int_{s=0}^{\infty} 
e^{-(\Ntot-\Npos+1)s} (1-e^{-s})^{\Npos}
\,\mathrm{d}s,\nonumber\\
\qquad&= \frac{1}{\Ntot+1}.
\label{eqn:ZGG}
\end{align}

At first sight, it may appear surprising that the marginal likelihood under the $\mathcal{G}\mathcal{G}$ prior is independent of $\Npos$. However, this behavior is a natural consequence of the defining property of the $\mathcal{G}$ prior: it is uniform over experimental \emph{outcomes}, rather than over the underlying Bernoulli probability itself.

The transformation $s=-\log(1-x)$ maps the prior $\mathcal{G}(x)\propto[-\log(1-x)]^{-1/2}$ into a flat measure over the cumulative hazard $s$, such that equal prior weight is assigned to each order of magnitude in the expected number of positive outcomes. When combined with a binomial likelihood, the marginalization over $x$ therefore averages uniformly over all possible outcome frequencies, rather than privileging any particular value of $\Npos/\Ntot$.

As a result, the evidence depends only on the total number of trials $\Ntot$, which sets the dimensionality of the outcome space, and not on the specific realization $\Npos$. In this sense, the $\mathcal{G}$ prior represents a state of maximal prior ignorance about the experimental outcome itself, and the observed $\Npos$-independence of $\mathcal{Z}_{\mathcal{G}\mathcal{G}}$ is thus an expected and diagnostic feature of this prior choice.

More challenging is the $f=0$ case, where we need to solve the integral:

\begin{align}
\mathcal{Z}_{f=0,\mathcal{G}\mathcal{G}} &= \int_{\CPP=0}^1
\mathcal{L}_{f=0} \mathcal{G}(\CPP)
\,\mathrm{d}\CPP,\nonumber\\
\qquad&= \int_{\CPP=0}^1 \frac{
\binom{\Ntot}{\Npos} (1-\CPP)^{\Ntot-\Npos} \CPP^{\Npos}
}{
\sqrt{\pi} \sqrt{-\log(1-\CPP)}
}\,\mathrm{d}\CPP.
\label{eqn:Zf0GG}
\end{align}

Although a closed-form solution can be expressed, it is in the form of a sum of finite alternating terms that becomes numerically unstable at large $\Npos$. We found a better approach was to re-write the integral as

\begin{align}
\mathcal{Z}_{f=0,\mathcal{G}\mathcal{G}} &= 
\frac{ 2 \binom{\Ntot}{\Npos} }{ \sqrt{\pi} }
\int_{t=0}^{\infty} e^{-(\Ntot-\Npos+1)t^2} (1-e^{-t^2})^{\Npos}\,\mathrm{d}t,
\end{align}

where we have used the substitution, $t = \sqrt{-\log\bar{\CPP}}$. The integral is then stable using a generalized Gauss-Laguerre scheme.

\subsubsection{Special case of $\Npos=0$}

As before, progress can be made by considering the special case of $\Npos=0$, since now the integral simplifies to

\begin{align}
\lim_{\Npos\to0} \mathcal{Z}_{f=0,\mathcal{G}\mathcal{G}} &= \int_{\CPP=0}^1 \frac{ (1-\CPP)^{\Ntot} }{ \sqrt{\pi} \sqrt{-\log(1-\CPP)} }\,\mathrm{d}\CPP,\nonumber\\
\qquad&= \frac{1}{\sqrt{\Ntot+1}}.
\label{eqn:Zf0GG_Npos0}
\end{align}

Combining with Equation~\ref{eqn:ZGG} yields a Bayes factor of

\begin{align}
\lim_{\Npos\to0} \mathcal{K}_{\mathcal{G}\mathcal{G}} &= \sqrt{\Ntot+1}.
\label{eqn:KGG_Npos0}
\end{align}

Thus, the evidence in favor of $f=0$ grows as the square root of the number of observations taken - in the case where all observations yield a null result.

\subsubsection{Special case of $\Npos=\Ntot$}

In the opposite limit of $\Npos=\Ntot$, the $f=0$ marginal likelihood (Equation~\ref{eqn:Zf0GG}) becomes

\begin{align}
\lim_{\Npos\to\Ntot} \mathcal{Z}_{f=0,\mathcal{G}\mathcal{G}} &= \int_{\bar{\CPP}=0}^1 \frac{ (1-\bar{\CPP})^{\Ntot} }{ \sqrt{\pi} \sqrt{-\log\bar{\CPP}} }\,\mathrm{d}\bar{\CPP}.
\end{align}

As with the general case, one could state a closed-form solution as a sum of finite alternating terms, but this becomes unstable at large $\Ntot$. Instead, we seek to approximate the integrand in the case of $\Ntot\gg1$ limit, since numerical experiments reveal this is where the critical threshold of strong evidence occurs. We first note that $\log(1-\bar{\CPP}) = -\bar{\CPP} - \bar{\CPP}^2/2 + ...$ and thus

\begin{align}
(1-\bar{\CPP})^{\Ntot} =& \exp\big(  -\bar{\CPP} - \bar{\CPP}^2/2 + ... \big),\nonumber\\
\qquad&\simeq e^{-\Ntot \bar{\CPP}},
\end{align}

since $\bar{\CPP}\in[0,1]$ and thus the higher order terms can be ignored. We can now extend the $\bar{\CPP}$ integral limits from $[0,1]$ to $[0,\infty]$ since $e^{-\Ntot\bar{\CPP}}$ will contribute negligibly when $\bar{\CPP}>1$ and $\Ntot\gg1$:

\begin{align}
\lim_{\Ntot\gg1} \lim_{\Npos\to\Ntot} \mathcal{Z}_{f=0,\mathcal{G}\mathcal{G}} &\simeq \int_{\bar{\CPP}=0}^{\infty} \frac{ e^{-\Ntot \bar{\CPP}} }{ \sqrt{\pi} \sqrt{-\log\bar{\CPP}} }\,\mathrm{d}\bar{\CPP}.
\end{align}

It is now convenient to make the substitution $q=\Ntot\bar{\CPP}$ (such that $\mathrm{d}y=\mathrm{d}q/\Ntot$), giving

\begin{align}
\lim_{\Ntot\gg1} \lim_{\Npos\to\Ntot} \mathcal{Z}_{f=0,\mathcal{G}\mathcal{G}} &\simeq \int_{q=0}^{\infty} \frac{ e^{-q} }{ \Ntot \sqrt{\pi} \sqrt{-\log(q/\Ntot)} }\,\mathrm{d}q,\nonumber\\
\qquad&\simeq \int_{q=0}^{\infty} \frac{ e^{-q} }{ \Ntot \sqrt{\pi} \sqrt{L-\log q } }\,\mathrm{d}q,\nonumber\\
\qquad&\simeq \int_{q=0}^{\infty} \frac{ e^{-q} }{ \Ntot \sqrt{\pi L} \sqrt{1-L^{-1}\log q } }\,\mathrm{d}q,
\end{align}

where we've used $L=\log \Ntot$. After rescaling to $q=\Ntot\bar{\CPP}$, the integrand contains $e^{-q}$, whic restricts the dominant contribution to $q\sim\mathcal{O}[1]$ as $\Ntot\to\infty$. In this region, $\log q \sim \mathcal{O}[1]$ and therefore $L^{-1}\log q=\log q/\log \Ntot$ is small. We may thus exploit the series expansion $(1-x)^{-1/2} = 1 + (x/2) + ...$ to re-write our integrand as

\begin{align}
\lim_{\Ntot\gg1} \lim_{\Npos\to\Ntot} \mathcal{Z}_{f=0,\mathcal{G}\mathcal{G}} &\simeq \int_{q=0}^{\infty}
\frac{ e^{-q} }{ \Ntot \sqrt{\pi L} } \Bigg( 1 + \frac{1}{2} \frac{\log q}{L} + \ldots \Bigg),\nonumber\\
\qquad&\simeq \frac{1}{\Ntot\sqrt{\pi\log\Ntot}},
\end{align}

where the last line only considers the leading order. Combining with Equation~\ref{eqn:ZGG} (and replacing $(\Ntot+1)\to\Ntot$ since $\Ntot\gg1$), the large-$\Ntot$ limiting Bayes factor asymptotically approaches

\begin{align}
\lim_{\Ntot\gg1} \lim_{\Npos\to\Ntot} \mathcal{K}_{\mathcal{G}\mathcal{G}} &\simeq \frac{ 1 }{ \sqrt{\pi\log\Ntot} }.
\label{eqn:KGG_NposA}
\end{align}

We may now solve the above for the critical threshold of strong evidence, which reveals that we require $\simeq 66.68$\,trillion measurements to establish the life hypothesis - all of which are positive detections. As before, this highlights the preposterous survey requirements to overwhelm a diffuse $\CPP$ prior.

\section{Special Case of a Truncated Confounder Prior}
\label{sec:truncated}

\subsection{Motivation}

Thus far, we have obtained two major results. First, if the confounding positive probability, $\CPP$, is a known fixed scalar (see Section~\ref{sec:knownCPP}) then cases where $\CPP\leq0.87$ permit for establishing the life hypothesis with $\Ntot=25$, although that limit corresponds to the fairly extreme cases of $\Npos=\Ntot$. Second, if $\CPP$ follows a diffuse prior (as the occurrence rate of life, $f$, is assumed to do so throughout), then $\Ntot=25$ is insufficient to establish either the null or affirmative hypothesis for all $\Npos$ - with the exception of a uniform prior on $\CPP$ and $\Npos=0$ or $\Npos=1$.

To complete the story, it is useful to consider somewhat of a hybrid between these two extremes where the confounder rate is a \textit{truncated} diffuse prior. For mathematical convenience, we consider a truncated uniform prior for $\CPP$ in what follows, where we keep the floor at zero but cap the ceiling to some sub-unity value e.g. $\CPP \sim \mathcal{U}[0,0.5]$. For consistency, we keep the $f$ prior uniform as well, but it is fully diffuse over the interval $[0,1]$ in all cases - since it's somewhat inconceivable that we could ever reasonably assert a truncation in either direction. The results obtained will allow us to smoothly connect the asymptotic behavior of varying the prior.

\subsection{General $\Npos$ case}

With this change, we can write the joint posterior distribution of $\{f,\CPP\}$ as:

\begin{align}
\underbrace{ \pdf(f,\CPP|\Ntot,\Npos) }_{=\mathcal{P}_{\mathcal{U}\mathcal{V}}} 
&= \frac{ \pdf(\Npos|\Ntot,f,\CPP) \pdf(f) \pdf(\CPP) }{ \underbrace{\pdf(\Npos|\Ntot)}_{=\mathcal{Z}_{\mathcal{U}\mathcal{V}}} },
\end{align}

where we add the subscript ``$\mathcal{U}\mathcal{V}$'' to denote this is for the case of a fully diffuse uniform prior on $f$ ($\mathcal{U}$) but a truncated uniform prior on $\CPP$ ($\mathcal{V}$). The marginal likelihood is found by double integration of the likelihood multiplied by the prior:

\begin{align}
\mathcal{Z}_{\mathcal{U}\mathcal{V}} &= \int_{\CPP=0}^{\Cmax} \int_{f=0}^1
\big( 1 - (1-f)(1-\CPP) \big)^{\Npos}
\big( (1-f)(1-\CPP) \big)^{\Ntot-\Npos}
\binom{\Ntot}{\Npos}\,\mathrm{d}f\,\mathrm{d}\CPP,
\end{align}

which evaluates to
\begin{align}
\mathcal{Z}_{\mathcal{U}\mathcal{V}} =& \frac{1}{\Cmax} \binom{\Ntot}{\Npos}
\Bigg( - ((\Ntot-\Npos)!)^2 (1-\Cmax)^{\Ntot-\Npos+1} \nonumber\\
\qquad& \times\, _3\tilde{F}_2[\Ntot-\Npos+1,\Ntot-\Npos+1,-\Npos;\Ntot-\Npos+2,\Ntot-\Npos+2;1-\Cmax]\nonumber\\
\qquad& + \log (1-\Cmax)
        \big(  B_{1-\Cmax}[\Ntot-\Npos+1,\Npos+1]
             + B_{\Cmax}[\Npos+1,\Ntot-\Npos+1] \nonumber\\
\qquad&      - B[\Npos+1,\Ntot-\Npos+1] \big)
	    + B[\Npos+1,\Ntot-\Npos+1] \big( H_{\Ntot+1}-H_{\Ntot-\Npos} \big) \Bigg),
\label{eqn:ZUV}
\end{align}

where $B_z[a,b]$ is the incomplete Beta function. In the limit of $\Cmax\to1$, we confirmed that this returns our earlier result of Equation~(\ref{eqn:ZUU}), as expected. In order to obtain the Bayes factor of hypothesis $\mathcal{H}_{f=0}$ or $\mathcal{H}_{f>0}$, we require the marginal likelihood in the limit of $f\to0$:

\begin{align}
\mathcal{Z}_{f=0,\mathcal{U}\mathcal{U}} &= \int_{\CPP=0}^{\Cmax}
\mathcal{L}_{f=0}
\,\mathrm{d}\CPP,\\
\qquad&= \frac{1}{\Cmax} \binom{\Ntot}{\Npos} B_{\Cmax}[\Npos+1,\Ntot+\Npos-1].
\label{eqn:Zf0UV}
\end{align}

Again, in the limit of $\Cmax\to1$ this returns our earlier result of Equation~(\ref{eqn:Zf0UU}), as expected. Accordingly, the Bayes factor is

\begin{align}
\mathcal{K}_{\mathcal{U}\mathcal{V}} =& \frac{ \mathcal{Z}_{f=0,\mathcal{U}\mathcal{V}} }{ \mathcal{Z}_{\mathcal{U}\mathcal{V}} }.
\label{eqn:KUV}
\end{align}

\subsection{Special case of $\Npos=0$}

The elaborate form of Equation~(\ref{eqn:ZUV}) makes it difficult to intuit the functional behaviour of the Bayes factor. We thus proceed, as we did before, to consider the extreme limits with respect to $\Npos$, starting with $\Npos=0$. Starting from Equation~(\ref{eqn:KUV}), we find

\begin{align}
\lim_{\Npos\to0} \mathcal{K}_{\mathcal{U}\mathcal{V}} &= \Ntot+1,
\label{eqn:KUV_Npos0}
\end{align}

which is identical to the result found with a double uniform prior with Equation~(\ref{eqn:KUU_Npos0}). This can be understood by the virtue of $\Npos=0$ corresponding to no confounders acting and the impact of $\Cmax$ dissolves. This result again implies that $\Ntot\geq9$ observations are needed for there to be any prospect of establishing the null hypothesis.

\subsection{Special case of $\Npos=\Ntot$}

In the opposite limit of $\Npos=\Ntot$, the Bayes factor (Equation~\ref{eqn:KUV}) becomes

\begin{align}
\lim_{\Npos\to\Ntot} \mathcal{K}_{\mathcal{U}\mathcal{V}} &= - \frac{ \Cmax^{\Ntot+1} }{
B_{\Cmax}[\Ntot+2,0] + \log(1-\Cmax)
}.
\label{eqn:KUV_NposA}
\end{align}

To reach the strong evidence threshold, we find the tail here is very sensitive to $\Cmax$. For example, if $\Cmax=0.9$, we get the encouraging result that $\Ntot\geq14$ would suffice. This result might seem puzzling since we know the double uniform prior result that if $\Cmax\to1$, the required number of observations explodes to $\Ntot=12366$ (and becomes trillions with less informative priors on $f$ and $\CPP$). The transition is indeed as steep as is implied. For example, with $\Cmax=0.99$ we have $\Ntot\geq83$, with $\Cmax=0.999$ we have $\Ntot\geq461$ and with $\Cmax=0.9999$ we have $\Ntot\geq2165$. Thus, even tiny adjustments to $\Cmax$ in this region have a huge influence on the Bayes factors and experimental design requirements. In practice, it is to difficult to imagine how one could assuredly argue that $\Cmax=0.999$ rather than $\Cmax=0.9999$, yet the impact is a nearly an order-of-magnitude increase in the required sample size. Indeed, this is why we suggest the most incontrovertible prior is completely diffuse over the interval $[0,1]$.

We illustrate the impact of varying $\Cmax$ on the Bayes factors, as well as comparing to the other priors considered in this work, in Table~\ref{tab:bayesfactors}.

\begin{table}
\caption{
Number of total observations, $\Ntot$, needed to reach two critical Bayes factors for an idealized experiment with an unknown occurrence rate, $f$, and confounding positive probability, $\CPP$. The third column corresponds to the minimum number of observations needed to strongly favor the null (no life) hypothesis, in the best case scenario of no positive detections ($\Npos=0$). The fourth column corresponds to the minimum number of observations needed to strongly favor the affirmative (life) hypothesis, in the best case scenario of all positive detections ($\Npos=\Ntot$).  We split the table into three panels corresponding to three types of priors investigated in this work. We argue the most agnostic and justifiable choice of prior is one of the double diffuse priors, for which the last row (the uniform-outcomes prior) gives arguably the most defensible selection.
} 
\centering 
\begin{tabular}{l l c c c} 
$\pdf(f)$ & $\pdf(\CPP)$ & \vline & 
$\mathcal{K}(f=0:f>0)=10|\Npos=0$ &
$\mathcal{K}(f=0:f>0)=0.1|\Npos=\Ntot$ \\ [0.5ex]
\hline
\multicolumn{5}{l}{\textbf{Fixed, known $\CPP$}}\\
\hline
$\mathcal{U}[0,1]$ & $\delta[0.00]$ & \vline & 9 & 1 \\
$\mathcal{U}[0,1]$ & $\delta[0.50]$ & \vline & 9 & 5 \\
$\mathcal{U}[0,1]$ & $\delta[0.75]$ & \vline & 9 & 13 \\
$\mathcal{U}[0,1]$ & $\delta[0.90]$ & \vline & 9 & 34 \\
$\mathcal{U}[0,1]$ & $\delta[0.99]$ & \vline & 9 & 359 \\
$\mathcal{U}[0,1]$ & $\delta[0.999]$ & \vline & 9 & 3613 \\
$\mathcal{U}[0,1]$ & $\delta[1.00]$ & \vline & 9 & $\infty$ \\
\hline
\multicolumn{5}{l}{\textbf{Truncated $\CPP$ prior}}\\
\hline
$\mathcal{U}[0,1]$ & $\mathcal{U}[0,0.00]$ & \vline & 9 & 1 \\
$\mathcal{U}[0,1]$ & $\mathcal{U}[0,0.50]$ & \vline & 9 & 3 \\
$\mathcal{U}[0,1]$ & $\mathcal{U}[0,0.75]$ & \vline & 9 & 6 \\
$\mathcal{U}[0,1]$ & $\mathcal{U}[0,0.90]$ & \vline & 9 & 14 \\
$\mathcal{U}[0,1]$ & $\mathcal{U}[0,0.99]$ & \vline & 9 & 83 \\
$\mathcal{U}[0,1]$ & $\mathcal{U}[0,0.999]$ & \vline & 9 & 461 \\
$\mathcal{U}[0,1]$ & $\mathcal{U}[0,1.00]$ & \vline & 9 & 12366 \\
\hline
\multicolumn{5}{l}{\textbf{Double diffuse priors}}\\
\hline
$\mathcal{U}[0,1]$ & $\mathcal{U}[0,1]$ & \vline & 9 & 12366 \\ 
$\mathcal{J}[0,1]$ & $\mathcal{J}[0,1]$ & \vline & 32 & ${\simeq}4.403 \times 10^{13}$ \\
$\mathcal{G}[0,1]$ & $\mathcal{G}[0,1]$ & \vline & 99 & ${\simeq}6.668 \times 10^{13}$ \\[0.5ex]
\hline
\end{tabular}
\label{tab:bayesfactors} 
\end{table}

\section{AB-Testing as a Possible Workaround}
\label{sec:divided}

\subsection{Motivation}

Thus far, we have seen that adopting fully diffuse priors for the prevalence of both life and confounders leads to extreme requirements for experiments seeking to detect life. \citet{foote:2023} proposed two pathways for life detection: i) a strong prior for life, or, ii) an unambiguous biosignature. As discussed in Section~\ref{sec:intro}, proposal i) minimizes the information gain from an experiment and is difficult to justify rigorously. Proposal ii) abandons the agnosticism we have adopted in this work. One might then wonder - is there a third way?

Confounders are not unique to life detection experiments - they exist in essentially all classification endeavors. Fortunately, there is well-understood and simple strategy to quantify their prevalence: control samples. Ideally, one would split the sample of $\Ntot$ measurements up into a control group and a test group. The control group would feature samples for which it was known with certainty that no effect was present and thus the confounder rate could be inferred from that clean set, allowing us to then interpret the test group. Of course, in astronomy, it's rarely so simple.

Consider, for example, attempting to design a control group that focussed on planets deemed to be likely inhospitable to life, such as Earths outside of the canonical habitable zone \citep{kopparapu:2013}. But a problem here is that the chemistry on such worlds would inevitably be distinct from the test sample (e.g. Earth analogs) and thus the confounder rate could not be justifiably treated as global across the two samples. Moreover, it is quite possible that planets outside of the canonical habitable zone (HZ) could sometimes host life too \citep{mol:2022}, thereby undermining the control labels.

Although a strict control group might not be feasible here, dividing the sample into two groups can still solve our dilemma. To avoid any pre-experiment prejudices, let's simply call the samples groups A and B. Each group has its own value for the prevalence of life, $f_A$ and $f_B$ - although it is certainly possible that $f_A=f_B$ too. The key assumption for this setup to work is that \textit{the confounder rate is the same for both}\footnote{
One might also imagine a hierarchical setup too, where the two populations have distinct $\CPP$ values but they are drawn from a global distribution. However, for the purposes of this paper we stick to the simpler non-hierarchical case, as our objective here is merely to show that a solution exists as a proof-of-principle.
}.

We'll discuss the realities of how this could plausibly manifest later in Section~\ref{sec:discussion}. For now, we adopt this scenario to merely highlight that a workaround does exist - as a proof-of-principle. Accordingly, we will not perform an exhaustive investigation of different priors as we did for a single population earlier, since even the single counter example described here will suffice to meet our outlined goal. Thus, in what follows, we adopt uniform priors on $f_A$, $f_B$ and $\CPP$.

\subsection{General $\{\NposA,\NposB\}$ case}

The likelihood function can now expressed as the probability of obtaining $\NposA$ positive labels from $\NtotA$ observations, and simultanesouly obtaining $\NposB$ positive labels from another $\NtotB$ observations. Thus the likelihood function is simply a product of two binomial likelihoods:

\begin{align}
\underbrace{\pdf(\NposA,\NposB|f_A,f_B,\CPP,\NtotA,\NtotB)}_{=\mathcal{L}_{AB}} =&
\big( 1 - (1-f_A)(1-\CPP) \big)^{\NposA}
\big( (1-f_A)(1-\CPP) \big)^{\NtotA-\NposA}
\binom{\NtotA}{\NposA}\nonumber\\
\qquad&\times \big( 1 - (1-f_B)(1-\CPP) \big)^{\NposB}
\big( (1-f_B)(1-\CPP) \big)^{\NtotB-\NposB}
\binom{\NtotB}{\NposB}.
\label{eqn:likelihoodAB}
\end{align}

With uniform priors, the joint posterior $\pdf(f_A,f_B,\CPP|\NposA,\NposB,\NtotA,\NtotB)$ is proportional to the likelihood. Now, the probability of obtaining $\NposA$ and $\NposB$ can be understood as the marginal likelihood (Bayesian evidence) of the data under the composite life model, found by integrating the following

\begin{align}
\mathcal{Z}_{\mathcal{U}\mathcal{U}\mathcal{U}} = \int_{\CPP=0}^1 \int_{f_B=0}^1 \int_{f_A=0}^1 \pdf(\NposA,\NposB|f_A,f_B,\CPP,\NtotA,\NtotB)\,\mathrm{d}f_A\,\mathrm{d}f_B\,\mathrm{d}\CPP,
\end{align}

where we add the ``$\mathcal{U}\mathcal{U}\mathcal{U}$'' to denote the triple uniform prior assumed here. The above can evaluated to

\begin{align}
\mathcal{Z}_{\mathcal{U}\mathcal{U}\mathcal{U}} = \binom{\NtotA}{\NposA} \binom{\NtotB}{\NposB} \sum_{i=0}^{\NposA} \sum_{j=0}^{\NposB} \frac{
(-1)^{i+j} \binom{\NposA}{i} \binom{\NposB}{j}
}{
(M_A+i)(M_B+j)(M_A+M_B-1+i+j)
},
\label{eqn:ZUUU}
\end{align}

where, for compactness, we define $M_A=\NtotA-\NposA+1$, $M_B=\NtotB-\NposB+1$. The case of $f=0$ is easier to tackle, being a single integral and evaluates to

\begin{align}
\mathcal{Z}_{f=0,\mathcal{U}\mathcal{U}\mathcal{U}} &= \int_{\CPP=0}^1  \Big( \lim_{f_B\to0} \lim_{f_A\to0} \mathcal{L}_{AB} \Big),\mathrm{d}\CPP,\nonumber\\
\qquad&= \binom{\NtotA}{\NposA} \binom{\NtotB}{\NposB} \frac{
(\NposA+\NposB)! (\NtotA+\NtotB-\NposA-\NposB)!
}{
\Gamma[\NtotA+\NtotB+2]
}.
\label{eqn:Zf0UUU}
\end{align}

For completion, the Bayes factor of obtaining the null hypothesis of $f_A=f_B=0$ (no life present) versus the affirmative hypotheses that includes life is

\begin{align}
\mathcal{K}_{\mathcal{U}\mathcal{U}\mathcal{U}} = \frac{
\mathcal{Z}_{f=0,\mathcal{U}\mathcal{U}\mathcal{U}}
}{
\mathcal{Z}_{\mathcal{U}\mathcal{U}\mathcal{U}}
}.
\label{eqn:KUUU}
\end{align}

We find that for a finite sample $\Ntot$ split into two samples, the number of $\{\NposA,\NposB\}$ combinations that lead to strong detections is consistently maximized when the $\Ntot$ is evenly split, for all $\Ntot$. This agrees with our intuition that splitting the sample into two should maximize the training data whilst minimizing the lost test data.

\subsection{Numerical Experiments with $\Ntot=24$}

To provide some further insight, we evaluate the Bayes factor in the HWO-like case of $\Ntot=24$ (we use 24 rather than 25 so that the sample can be split evenly), split into two even groups. We vary $\NposA$ and $\NposB$ across the 2D grid of all possible integer outcomes and evaluate the Bayes factor each time.

\begin{figure*}
\begin{center}
\includegraphics[width=13.0cm,angle=0,clip=true]{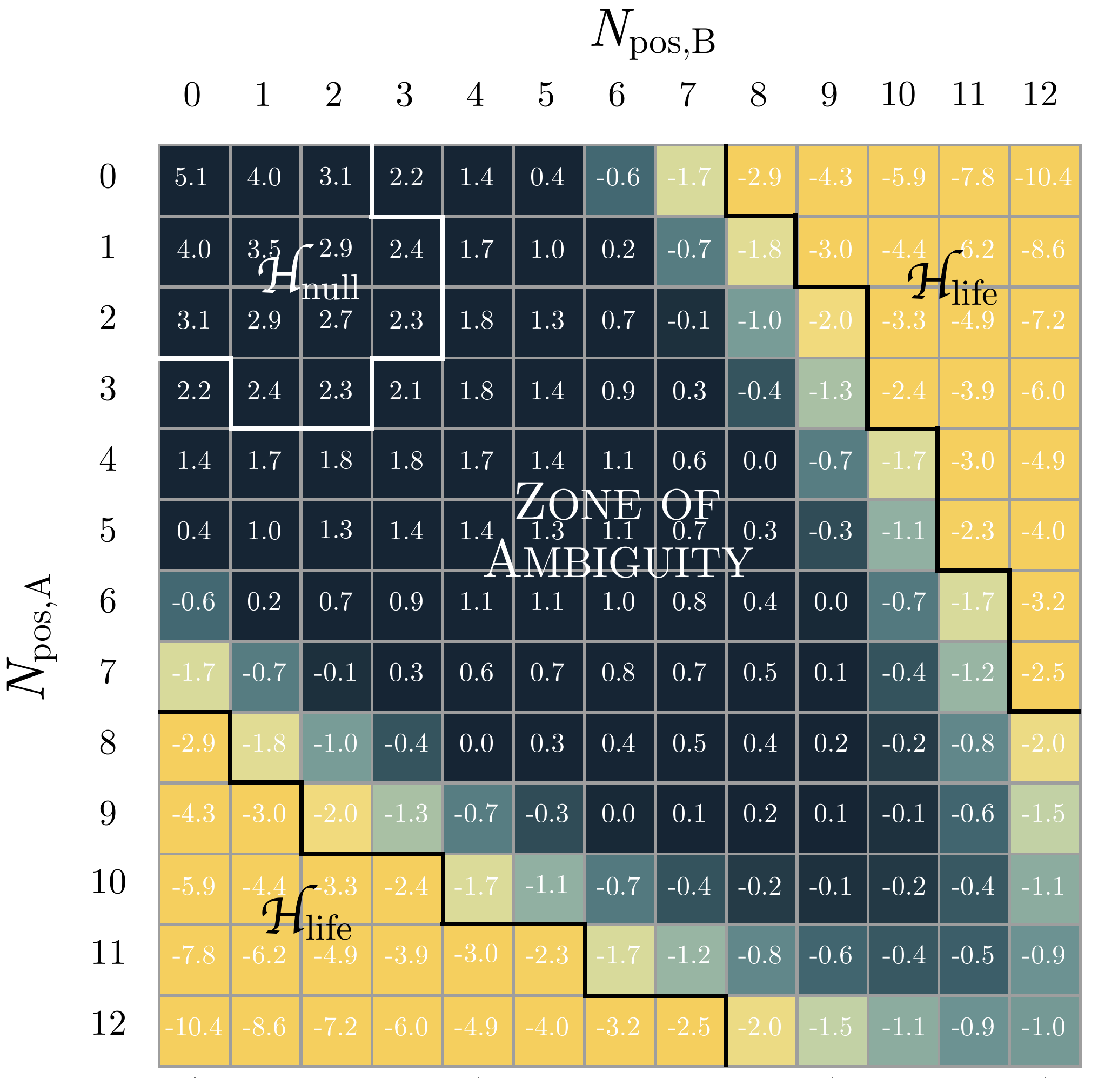}
\caption{
Grid plot of the log-Bayes factor for the null (no life) hypothesis versus the affirmative (life) hypothesis ($\mathcal{K}_{f0}$) when a sample of $\Ntot=24$ targets are split into two groups with differing life rates but a global confounder rate. Although a large zone of ambiguity persists, the life hypothesis can be strongly claimed in 40 of 169 possible outcomes. For comparison, with a monolithic population we find that no outcomes could ever achieve this.
}
\label{fig:doublegrid}
\end{center}
\end{figure*}

In Figure~\ref{fig:doublegrid}, we illustrate the results in a matrix plot, which reveals that the zone of ambiguous results spans most outcomes, but cases with $|\NposA-\NposB|\geq7$ can (but not always) produce strong life detections. This can be understood as the result of one group having far more detections than the other despite sharing the same confounder rate, thereby demanding distinct values of $f_A$ and $f_B$, and ruling out the possibility that both vanish.

Figure~\ref{fig:doublegrid} also reveals how cases where the null (no life) hypotheses are favored cluster in the corner of low $\{\NposA,\NposB\}$, which mirrors what we found earlier in the single population case (e.g. see Figure~\ref{fig:diffuse}).

\begin{figure*}
\begin{center}
\includegraphics[width=16.0cm,angle=0,clip=true]{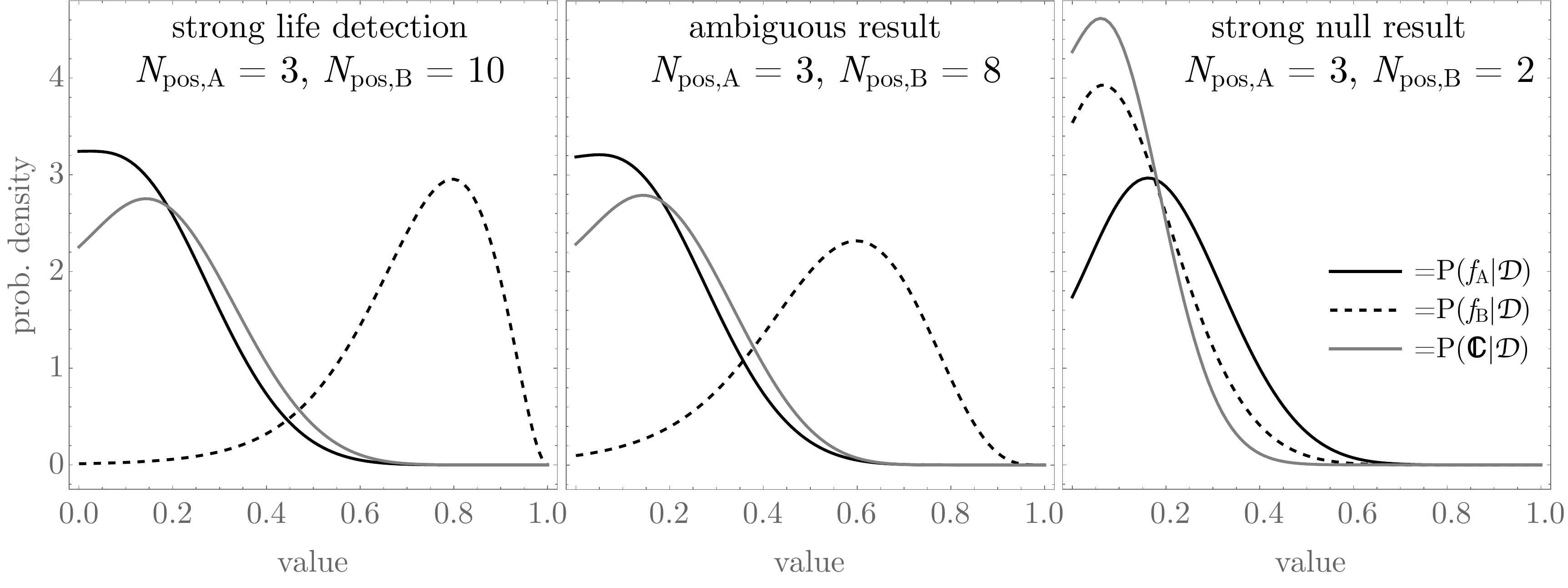}
\caption{
Three examples of the marginalized posteriors obtained for $f_A$, $f_B$ and $\CPP$ when $\Ntot=24$ and is divided into two groups, A and B. Each panel corresponds to a grid position show in Figure~\ref{fig:doublegrid}, with the data scenarios given by the panel heading. From left to right, we slide from a strong life detection to a strong null detection.
}
\label{fig:doubleposts}
\end{center}
\end{figure*}

As a more detailed breakdown of this grid of results, Figure~\ref{fig:doubleposts} shows the marginalized posterior distributions for $f_A$, $f_B$ and $\CPP$ obtained for three selections amongst our grid. These three show the effect of sliding from strong preference $\mathcal{H}_{\mathrm{life}}$ to an ambiguous result and then to strong preference for $\mathcal{H}_{\mathrm{null}}$.

The most encouraging takeaway from Figure~\ref{fig:doublegrid} is that 40 of 169 ($(1+\Ntot/2)^2$) possible outcomes (24\%) indicate strong evidence for the life hypothesis - even with diffuse priors. In contrast, with a monolithic population and double-uniform priors, there wasn't a single outcome that could do this. If we define a ``definitive'' experimental outcome as one for which strong evidence is obtained for at least one of two hypotheses (i.e. not an ambiguous outcome), then 61 of the outcomes are definitive (36\%).

\subsection{Varying $\Ntot$}

\begin{figure*}
\begin{center}
\includegraphics[width=12.0cm,angle=0,clip=true]{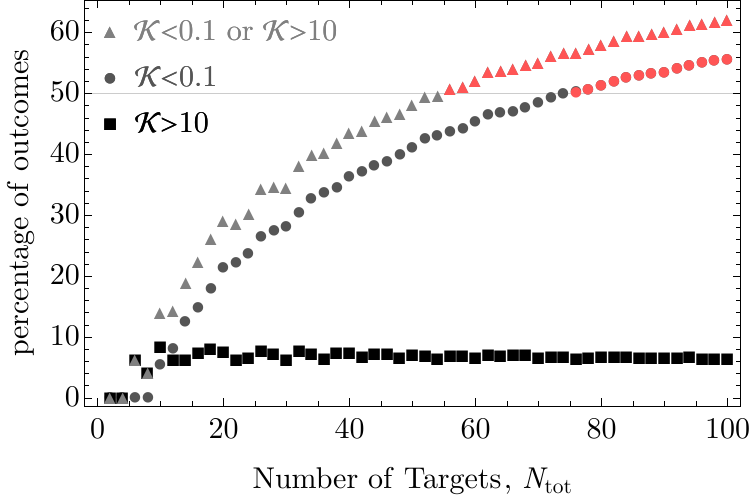}
\caption{
Assuming an evenly-divided AB sample of $\Ntot$ targets, we show the fraction of experimental outcomes that lead to strong evidence in favor of one of the two hypotheses, the life- and null-hypothesis. The fraction of outcomes that favor a null result converges to $\sim6.3$\%, whereas the pathway to a life detection grows with $\Ntot$, exceeding half of samples (horizontal line) for $\Ntot\geq76$ (red circles). For either case to be strongly favored, this becomes $\Ntot\geq56$ (red triangles).
}
\label{fig:definitives}
\end{center}
\end{figure*}

Through numerical experimentation shown in Figure~\ref{fig:definitives}, we found that setting $\Ntot\geq76$, divided into two sets, would cause the the majority of outcomes to yield strong detections for the life hypothesis. For a strong detection for either hypothesis, this becomes $\Ntot\geq56$.

This is promising and potentially within scope of a slightly more ambitious HWO survey \citep{feinberg:2024,stark:2024}. However, we caution that we were unable to find a universal prior for $f_A$, $f_B$ and $\CPP$ that produces uniform-outcomes across the $\{\NposA,\NposB\}$ grid (unlike Section~\ref{sub:GG}). Indeed, our analysis suggests that no such prior exists under a shared-$\CPP$ model. As a result, the triple uniform prior here produces a heterogeneous weighting of each outcome. Regardless, the adoption of uniform priors for variables such as this is a very common practice and thus we consider these results crucial in guiding future surveys for life.

\subsection{Impotent AB-Testing}
\label{sub:impotent}

Figure~\ref{fig:doublegrid} reveals that even with the AB-testing approach, a definitive result is not guarenteed - a wide zone of ambiguity exists. Crucially, inspection of Figure~\ref{fig:doublegrid} reveals that this zone lies along the diagonal of $\NposA \sim \NposB$, corresponding to $f_A \sim f_B$. Indeed, we should naturally expect that if we split the sample into two with identical $f$ and $\CPP$ values, then the split is completely arbitrary and no new information can be gleaned.

This can be formally demonstrated by revisiting our likelihood function and formally setting $\NtotA = \NtotB = \Ntot/2$ and $f_A=f_B=f$ yielding

\begin{align}
\lim_{f_A=f_B=f} \lim_{\NtotA=\NtotB=\Ntot/2} \mathcal{L}_{AB} &\propto
(1-(1-f)(1-\CPP))^{\NposA+\NposB} ((1-f)(1-\CPP))^{\Ntot-\NposA-\NposB},
\label{eqn:impotent}
\end{align}

which is the same likelihood found earlier with a monolithic population in Equation~(\ref{eqn:likelihood}). This reveals a crucial (but intuitive) warning for AB-testing - the sample can't be arbitrarily split. Groups A and B need different $f$ values for this approach to work, and the greater the difference then the better our leverage.

\section{Discussion}
\label{sec:discussion}

\subsection{Catastrophic Agnosticism}

Even for an idealized experiment with perfect sensitivity and recall, a positive detection label of some signature of life can be obtained via two mechanisms: 1) the genuine presence of life, or, 2) a confounder. The confounder positive probability, $\CPP$, thus represents the primary threat to a convincing life detection, as seen with the numerous historical claims of life that failed to obtain scientific consensus. 

Certainly, setting $\CPP=0$ maximizes the propensity of a survey to make a positive claim - since any positive detection labels automatically require a non-zero prevalence of life, $f>0$. As we showed in Section~\ref{sec:knownCPP}, one can go further and set $\CPP$ to almost any fixed value of our choosing and still retain a clear pathway to life detection. But the principal challenge facing confounders is an epistemic one - we don't know what we don't know. Accordingly, the $\CPP$ could be anything.

A similar challenge faces the prevalence rate of life, for which a default assumption is that we have no a-priori prejudice for how common or rare life should be (e.g see \citealt{angerhausen:2025,finetuning:2025}). This position of agnosticism is reflected by the appropriate use of diffuse (uninformative) priors on said rate, $f$. Since a positive detection label can be equally obtained via life or confounders, and we are forced to concede epistemic shortcomings for both terms, it is difficult to justify a non-diffuse prior for $\CPP$ as well.

As shown in this work, specifically in Section~\ref{sec:diffuse}, adopting diffuse priors for both $f$ and $\CPP$ is catastrophic for the prospects of life, especially with HWO-scale surveys. The number of targets required certainly exceeds $10^4$, and is arguably more like $10^{13}$. Such a large sample is far beyond the scope of near-term biosignature searches \citep{feinberg:2024}, although we note it is plausibly within reach of some technosignature searches \citep{suazo:2024}. However, even here the life hypothesis can only overwhelm the confounder hypothesis if the number of detections is extraordinarily large, greatly blunting the sensitivity of such surveys.

\subsection{AB Testing}

We have proposed a workaround that exploits AB-testing; dividing the sample into two groups, for which an even split produces maximal sensitivity. This approach allows for each subset to have unique $f$ value (a strict control group with $f=0$ is not required), but it formally demands a global confounder rate. With this approach, even a modest HWO-like sample of $\Ntot=24$ would produce strong life detections in one-quarter of possible outcomes. Tripling the sample size ($\Ntot\geq76$) doubles the number of outcomes leading to strong life detections. This approach is attractive since it retains full agnosticism about life and confounders, and can delivery useful constraints without greatly inflating the sample size.

However, we concede here a potentially major obstacle to realizing even this approach. If we split a sample into two, any cut for which $f_A \sim f_B$ nullifies the effectiveness of this workaround (as shown in Section~\ref{sub:impotent}). In general, arbitrarily splitting the sample in two will indeed satisfy $f_A \sim f_B$. Thus, the split has to be carefully chosen in such a way that we a-priori expect $|f_A - f_B|>0$; and, the larger this difference the greater leverage the survey will have. But therein lies the problem - if we divide the sample into two groups based on some astrophysical criteria (in order to engineer $|f_A - f_B|>0$), would not that same process undermine the crucial assumption that the confounder rate is the same?

We strongly encourage the community to consider possible viable ways in which such AB-testing could be performed where one could reasonably justify that $f_A \neq f_B$ but $\CPP_A = \CPP_B$. An agnostic analysis of any near-term biosignature search will hinge upon this point. We tentatively suggest that one such split could be A) Earth-mass planets in the HZ of their stars but with no other such planets in the same region, and, B) Earth-mass planets that reside in HZ of their stars along with one or more other such planets also in the HZ. If panspermia is effective between neighboring Earth analogs, then one should reasonably expect $f_B > f_A$, boosted by this effect. In contrast, there are no obvious physical differences between these planet groups that would drive different confounder mechanisms.

Unfortunately, we do not yet know the occurrence rate of multiple Earths in the HZ of HWO target stars (or arguably even single Earths). It may so low that the target stars would be too faint for HWO to ever realistically pursue this strategy. Yet more, there are numerous ways in which this simple proposal is torpedoed. Figure~\ref{fig:earths} illustrates six examples, where a class of seeming one/multi-HZ planets is in fact not what it appears. Perhaps the most serious of these is the calculation of the HZ, and indeed even the very notion that a strict boundary exists at all - it is plausible planets far outside this zone could sustain life through a variety of mechanisms.

We finish by briefly discussing the mirror case where $f_A = f_B$ but $\CPP_A \neq \CPP_B$. Since both pathways are symmetrically equivalent ways of generating life signals, the mathematics is identical to the AB testing scheme outlined with $f_A \neq f_B$. Naively, this seems like an even better approach. One could imagine surveying the same sample of planets for two different biosignatures, A and B. If we define $f$ as the fraction of cases where both biosignatures manifest to detectable levels, then $f_A = f_B$ by construction, and the $\CPP$ of each could trivially be different. However, this framing has made the mistake of slipping into the original thinking that motivates the entire paper - a biosignature detection (or even a pair of them) is not equivalent to the detection of life. There could be unknown confounders which generate both. All this experimental setup could hope to achieve is establishing that both biosignatures are present. It cannot directly address the life question. Put another way, life (at least as considered throughout this work) is a binary label that can't be trivially split into two sub-types like this.

These thought experiments raise serious concerns about how an agnostic search for life could ever hope to statistically establish its existence, especially with the relatively information poor regime of seeking molecular fingerprints of life. But, in principle, pathways do exist and demand deep thought about possible avenues forward. This author is confident these challenges can be overcome, that someone will find a clever framing to dissolve these issues, even if the answer alludes this author at the time of writing.

\appendix
\section{Derivation of $\mathcal{K}_{\mathcal{J}\mathcal{J}}$ for $\Npos=\Ntot$ and $\Ntot\gg1$}
If we define $p = 1 - (1-f)(1-\CPP)$, then the double integral is a weighted expectation value of $p^{\Ntot}$, and thus in the limit of $\Ntot\gg1$ only $p\sim1$ values will meaningfully contribute. If $q=1-p$, then $E[(1-q)^{\Ntot}] \simeq E[e^{-\Ntot q}]$. So our problem simplifies to a Laplace-type average over small $q$. Writing $\bar{f}=1-f$ and $\bar{\CPP}=1-\CPP$, $\bar{f}$ and $\bar{\CPP}$ will follow the same Jeffrey's prior as $f$ and $\CPP$ due to the symmetry of the distribution. Since only values $p\sim1$ contribute, we only need consider $q\sim0$ corresponding to $\bar{f}\sim0$ and $\bar{\CPP}\sim0$, where the prior looks like $\pdf(x) \simeq 1/(\pi\sqrt{x})$. The distribution of $q$ will follow

\begin{align}
\pdf(q) &= \int_{q}^{1} \pdf(x) \pdf(q/x) (1/x) \mathrm{d}x,\nonumber\\
\qquad&= \frac{\log(1/q)}{\pi^2 \sqrt{q}}.
\end{align}

Now, given its nature as an expectation value, the integral can be expressed as

\begin{align}
\lim_{\Ntot\gg1} \lim_{\Npos\to\Ntot} \mathcal{Z}_{\mathcal{J}\mathcal{J}} &\simeq 
\int_{0}^1 
e^{-\Ntot q} \pdf(q)\,\mathrm{d}q,\nonumber\\
\qquad&= \frac{1}{\pi^2\sqrt{\Ntot}} \int_{q=0}^{\Ntot} e^{-t} t^{-1/2} \log(n/t)\,\mathrm{d}t,
\end{align}

where the second line makes the substitution $t=q \Ntot$. Because of the $e^{-t}$ sharp drop off at high $t$, the upper integral limit can be set to $\infty$ without perturbing the result much. Further, since $\log(\Ntot/t) = \log\Ntot - \log t$ and $t=q \Ntot$ where $q\sim0$, then $\log(\Ntot/t) \simeq \log\Ntot$. The integral now simplifies to

\begin{align}
\lim_{\Ntot\gg1} \lim_{\Npos\to\Ntot} \mathcal{Z}_{\mathcal{J}\mathcal{J}} &\simeq 
\frac{\log \Ntot}{\pi^2\sqrt{\Ntot}} \int_{q=0}^{\infty} e^{-t} t^{-1/2}\,\mathrm{d}t,\nonumber\\
\qquad&= \frac{\log \Ntot}{\pi^2\sqrt{\Ntot}} \sqrt{\pi}.
\end{align}

Finally, we need $\mathcal{Z}_{f=0,\mathcal{J}\mathcal{J}}$ in the same limit. Starting from Equation~\ref{eqn:Zf0JJ}:

\begin{align}
\lim_{\Npos\to\Ntot }\mathcal{Z}_{f=0,\mathcal{J}\mathcal{J}} &= \frac{
\Gamma[ \Ntot+\tfrac{1}{2} ]
}{
\sqrt{\pi} \Ntot!
}.
\label{eqn:Zf0JJ_NposA}
\end{align}

In the limit of large $\Ntot$, this becomes

\begin{align}
\lim_{\Ntot\gg1} \lim_{\Npos\to\Ntot }\mathcal{Z}_{f=0,\mathcal{J}\mathcal{J}} &\simeq \frac{
1
}{
\sqrt{\pi \Ntot}
},
\end{align}

and thus the Bayes factor of interest is given by

\begin{align}
\lim_{\Ntot\gg1} \lim_{\Npos\to\Ntot} \mathcal{K}_{\mathcal{J}\mathcal{J}} &\simeq \frac{\pi}{\log\Ntot}.
\end{align}

\section{Derivation of the ``Uniform-Outcomes'' Prior}
Since $\Npos$ is drawn from a binomial distribution, imposing a uniform distribution for $\Npos$ is equivalent to demanding a uniform distribution on $p=1-(1-f)(1-\CPP) = 1-\bar{f}\bar{\CPP}$. This can be demonstrated by considering the marginalized probability of obtaining $\Npos$ successes from $\Ntot$ trials, which will be

\begin{align}
\pdf(\Npos|\Ntot) &= \int_{p=0}^1 \binom{\Ntot}{\Npos} p^{\Npos} (1-p)^{\Ntot-\Npos}\,\mathrm{d}p,\nonumber\\
\qquad&= \binom{\Ntot}{\Npos} B[\Npos+1,\Ntot-\Npos+1],\nonumber\\
\qquad&= \binom{\Ntot}{\Npos} \frac{\Npos!(\Ntot-\Npos)!}{(\Ntot+1)!},\nonumber\\
\qquad&= \frac{1}{\Ntot+1},
\end{align}

and thus $\Npos$ is uniform on $\{0,...,\Ntot\}$. A uniform distribution in $p$ is equivalent to a uniform distribution in the product $\bar{f}\bar{\CPP}$, due to symmetry. This product becomes a sum in log-space, such that we can define $Y_f = -\log\bar{f}$ and $Y_C = -\log\bar{\CPP}$ and thus $\bar{f}=e^{-Y_f}$ and $\bar{\CPP} = e^{-Y_C}$ where the $Y$ variables span $[0,\infty]$. We now have

\begin{align}
-\log(\bar{f}\bar{\CPP}) &= -\log\bar{f} - \log\bar{\CPP} = Y_f + Y_C.
\end{align}

If we choose $Y_{f,C} \sim \mathrm{Gamma}[a,1]$ (a Gamma distribution), where $a$ is a shape parameter, we can exploit the fact that the sum of two independent Gamma variables is another Gamma distribution where the shape equals the sum of shapes i.e. $Y_f + Y_C \sim \mathrm{Gamma}[2a,1]$. By choosing $a=1$, this is simply $\mathrm{Gamma}[1,1] \equiv \mathrm{Exp}[1]$. This serves our goal of obtaining a uniform distribution in $p$, since if $Y_f+Y_C \sim \mathrm{Exp}[1]$ then $e^{Y_f+Y_C} \sim \mathcal{U}[0,1]$.

We have thus established that setting $-\log\bar{f} = Y_f \sim \mathrm{Gamma}[\tfrac{1}{2},1]$ (and similarly for $\bar{\CPP}$) produces a uniform distribution in $p$. Thus, we have

\begin{align}
\pdf(Y_f) &= \frac{1}{\Gamma[\tfrac{1}{2}]} Y_f^{-1/2} e^{-Y_f}.
\end{align}

Since $f = 1 - e^{-Y_f}$, we can obtain the corresponding distribution in $f$ by change of variables, such that $Y_f = -\log(1-f)$ and $\mathrm{d}Y_f/\mathrm{d}f = 1/(1-f)$, giving

\begin{align}
\pdf(f) &= \frac{1}{\sqrt{\pi}} (-\log(1-f))^{-1/2} (1-f) \frac{1}{1-f},
\end{align}

and similarly for $\CPP$. Simplifying and casting as a general prior, one can write

\begin{align}
\mathcal{G}(x) &= \frac{1}{ \sqrt{\pi} \sqrt{-\log(1-x)} }.
\end{align}

\section*{Acknowledgements}

Special thanks to donors to the Cool Worlds Lab, without whom this kind of research would not be possible:
Philip Johnston, Gerrit Thomsen, Brian Cartmell, Mike Hedlund, Tom Donkin, Bas Gaalen, Emerson Garland, Axel Nimmerjahn, Brad Bueche, Chad Souter, Craig Frederick, Douglas Daughaday, Drew Aron, Ieuan Williams, Jason Rockett, Josh Alley, Mark Elliott, Mathew Farabee, Philip Johnston, Ryan Provost, Steve Larter, Tristan Zajonc, Warren Smith, Guillaume Saint, Marisol Adler, Leigh Deacon, Alex Vaal, Andrew Schoen, Benjamin Kingston, Hunter Schiff, Jason Bryant, John Morrison, Marcus Gillette, Nicholas Haan, Paul Borisoff, Richard Williams, Stephen Lee, Steven Patterson, Terriss Ford, Trevor Edris \& Zachary Danielson.

\bibliography{manuscript}{}
\bibliographystyle{aasjournalv7_surnames_only}



\end{document}